# Time-reversal-like degeneracies distinguished by the anomalous Hall effect in a metallic kagome ice compound

K. Zhao[1, 2, *], Y. Tokiwa[1], H. Chen[3, 4, *], and P. Gegenwart[1*]

1. *Experimentalphysik VI, Center for Electronic Correlations and Magnetism, University of Augsburg, 86159 Augsburg, Germany*

2. *School of Physics, Beihang University, Beijing 100191, China*

3. *Department of Physics, Colorado State University, Fort Collins, CO 80523-1875, USA*

4. *School of Advanced Materials Discovery, Fort Collins, CO 80523-1875, USA*

*Corresponding author: kan_zhao@buaa.edu.cn, huachen@colostate.edu, and philipp.gegenwart@physik.uni-augsburg.de

In magnetic crystals, despite the explicit breaking of time-reversal symmetry, two equilibrium states related by time reversal are always energetically degenerate. In ferromagnets such time-reversal degeneracy can be manifested by hysteresis loops in the magnetic field dependence of the magnetization and, if metallic, in the anomalous Hall effect. Importantly, both quantities simply change signs but not their absolute sizes under time reversal, which follows from their fundamental definitions. Our integral experimental and theoretical study shows that in the metallic kagome spin ice HoAgGe subject to finite magnetic fields parallel to the kagome plane, an emergent time-reversal-like degeneracy appears between magnetic states that have the same energy and net magnetization, but different sizes of the anomalous Hall effect. These degeneracies are unraveled by finite hysteresis in the field-dependent anomalous Hall effect contrasted with the vanishing hysteresis in the magnetization, which appears only at low-temperatures T<4K when the kagome spin ice is fully ordered into the $\sqrt{3} \times \sqrt{3}$ state. By explicitly determining the degenerate states and calculating the corresponding physical properties using a tight-binding model, we nailed down the time-reversal-like operation that transforms these degenerate states into each other. The operation is related to the nontrivial distortion of the kagome lattice in HoAgGe and is effective only because of the richness of degenerate states unique to kagome spin ice. Our work points to the powerful role of the anomalous Hall effect to diagnose hidden symmetries in frustrated spin systems.

Geometrically frustrated spin systems are a promising platform for novel states of matter[1-7]. Although magnetic neutron diffraction and magnetometry measurements are direct probes on the spin sector[6, 7], when the systems of study are sufficiently conductive, electric transport can often provide an enlightening perspective toward understanding exotic spin ordering and its consequences[8-12]. This is especially the case when the transport coefficients can be directly related to symmetries of the spin order[13-23]. A well-known example is the anomalous Hall effect (AHE) observed in pyrochlore iridate $Pr_2Ir_2O_7$ which provides evidence of time-reversal-symmetry breaking despite the absence of long-range magnetic ordering[10].

As a fundamental discrete symmetry operation, time reversal $\mathcal{T}$ plays a particularly important role in magnetic crystals. Adding $\mathcal{T}$ to space group operations leads to the establishment of magnetic space groups which allow systematic classification and investigation of magnetic phases both experimentally and theoretically. Moreover, governed by basic principles of quantum mechanics, almost all measurable physical quantities have definite transformation properties under $\mathcal{T}$, which can often provide system-independent predictions. For example, for magnetic phases that break the $\mathcal{T}$ symmetry, two states related by $\mathcal{T}$ are energetically degenerate (a macroscopic manifestation of Kramers theorem). If the magnetic phase is characterized by a nonzero net magnetization $\mathbf{M}$, the two phases must have mutually antiparallel $\mathbf{M}$ of same size and if the phase displays the AHE[8, 9], the AHE must also have opposite signs but same sizes in the two $\mathcal{T}$-related states, a consequence of Onsager reciprocity[24-26]. Recently it has been demonstrated both theoretically and experimentally that in certain noncollinear antiferromagnets the AHE can distinguish $\mathcal{T}$-symmetry breaking even though $\mathbf{M}$ is vanishing, although the AHE for the two $\mathcal{T}$-related states still has the same size[13-23]. In this work, however, we show that it is possible to use the AHE to distinguish degenerate states related by an emergent discrete symmetry operation $\mathcal{X}$ in a metallic kagome spin ice[27-37] compound HoAgGe under finite magnetic fields parallel to the kagome plane. Different from $\mathcal{T}$, $\mathbf{M}$ is unchanged under $\mathcal{X}$, while the AHE changes its *size* (Fig. 1a).

HoAgGe is an intermetallic compound consisting of stacked kagome lattices with nontrivial distortion formed by $Ho^{3+}$ ions[38]. Considering the $Ho^{3+}$ ions alone, the distortion reduces the $D_{6h}$ symmetry of the kagome lattice to $D_{3h}$, but does not change the local easy axes of the $Ho^{3+}$ moments. The strong local easy-axis anisotropy together with ferromagnetic nearest-neighbour coupling of the $Ho^{3+}$ moments lead to "1-in-2-out" or "2-in-1-out" ice rules on the kagome lattice[29-34]; the massive degeneracy resulting from the ice rules is weakly lifted by further-neighbour interactions of the RKKY type, leading to a $\sqrt{3} \times \sqrt{3}$ ground state of the dipolar kagome ice below 4K through a partially-ordered phase with decreasing temperature[38]. Interestingly, at low temperatures the magnetization versus external magnetic fields parallel to the kagome plane exhibits a series of plateaus, due to the competition between the $Ho^{3+}$ moments' coupling with the fields and the further-neighbour interactions among them[38]. In this work we show that, partly due to the nontrivial distortion of the kagome lattice in HoAgGe[39,40], a time-reversal-like $\mathscr{X}$ symmetry operation as introduced above becomes effective on the magnetic plateaus.

## Contrasting hysteretic behavior of the AHE and magnetization

HoAgGe crystals were cut into different shapes for magnetic and transport measurements with the demagnetization correction considered. As shown in Fig. S1[41], the demagnetization correction enables quantitative comparison between different measurements on differently shaped crystals. Fig. 1b displays a phase diagram of HoAgGe under $\mathbf{H}//b$ based on magnetic susceptibility ($\chi$), magnetic specific heat ($C_{mag}$), and magnetic Grüneisen parameter ($\varGamma_{mag}$) data. The color coding in Fig. 1b is based on low-temperature magnetic susceptibility (Fig. S3). Meta-magnetic transitions into the two lower-field states labelled by 1/3 and 2/3 plateaus first appear below ~7K (Fig. S2), where the fraction means the ratio between the net magnetization on a given plateau and that in the saturated state. These values can be understood by enforcing the kagome ice rules in a $\sqrt{3} \times \sqrt{3}$ unit cell. A 1/6 plateau, whose magnetic configuration is still unclear at the moment, is observed around 4K (Fig. S4). Between 1T and 2T, the sharp peak in $C_{mag}$ around 10K marks the transition from the paramagnetic phase to a partial-ordered phase presumably similar to that under zero field, while the broad peak around 7K corresponds to the transition into a low-temperature metastable phase (1/3 plateau) with Am'm2' symmetry (Fig. 1a and Figs. S5c, d). Above 2T, the sharp peak in $C_{mag}$ and $\varGamma_{mag}$ around 8K marks the transition from the paramagnetic state to another metastable state (2/3 plateau) with Am'm2' symmetry (Fig. S5e). The intermediate region between the two phase boundaries is expected to consist of both P-6'm2' and Am'm2' components. The metamagnetic transition into the saturated state is accompanied by the disappearance of the magnetic vector (1/3, 1/3, 0) in neutron data, with the magnetic unit cell becoming the same as the structural one. (See Fig. S4 for other thermodynamic data.)

Fig. 2 summarizes field- and temperature-dependent magneto-transport data with $\mathbf{H}//b$. The Hall resistivity $\rho_{ca}$ is measured with currents along $c$ and voltage along $a$. When $H<3.2$ T (dashed line in Fig. 2a), both the Hall resistivity and the magnetoresistance (MR) show clear changes at the metamagnetic transitions. The MR also has a pronounced jump at the 1/6 plateau at 2 K. At 2 K, $\rho_{ca}$ overall increases with the field despite the sudden changes at the metamagnetic transition. At 3T when the system transitions to the fully-polarized state, $\rho_{ca}$ changes rather abruptly with its sign also being reversed. Such a feature becomes less pronounced and shifts to lower fields with increasing temperature and eventually disappears at ~11 K, in agreement with Fig. 1b. Previous density functional theory calculations (see Fig. S20 of Ref. 38) indicate that HoAgGe has multiple bands crossing the Fermi energy in the paramagnetic state, which is consistent with a nonlinear-in-$H$ Hall resistivity at 50 K, shown in Fig. S9a. The sign change of $\rho_{ca}$ around 3T shown in Fig. 2 may therefore be related to significant changes of the band structure near the Fermi energy driven by the change of the magnetic order.

More interestingly, Fig. 2 shows that at 2K, the magnetotransport data have clear hysteresis for the 1/3 and 2/3 plateaus, in contrast to the indiscernible hysteresis in the magnetization versus field curves. In the middle of each plateau the physical observables (magnetization, transport coefficients, etc.) have weak dependences on magnetic fields, suggesting that the magnetic order is sufficiently uniform. The hysteretic behavior of $\rho_{ca}$, together with the same critical field values of the metamagnetic transitions at two halves of the hysteresis loop, strongly indicate the existence of at least two degenerate states with the same magnetization on each plateau when $H$ is between 1T and 3T.

To show that the hysteresis of $\rho_{ca}$ is inseparable from the finite-magnetic-field ordering derived from the $\sqrt{3} \times \sqrt{3}$ ground state, we directly compare the thermodynamic and transport data versus $\mathbf{H}$ along $b$ at two temperatures (8K and 3K) in Fig. 3. At 8K (Fig. 3a), when there is only one metamagnetic transition to the polarized state, there is one sharp peak in magnetic specific heat, whose position agrees with the zero-crossing of $\Gamma_{\mathrm{mag}}$, due to the entropy accumulation at the phase boundary between the partial-ordered phase and the fully saturated phase. Indeed, the calculated magnetic entropy displays a sharp peak and decreases upon further raising magnetic field. The $\rho_{ca}$ data at 8K display no hysteresis. At 3K (Fig. 3b) when the $\sqrt{3} \times \sqrt{3}$ ground state is realized at zero field, the three metamagnetic transitions are clearly observed in $C_{\mathrm{mag}}$, $\Gamma_{\mathrm{mag}}$, and $S_{\mathrm{mag}}$.

**Time-reversal-like degenerate states with different AHE sizes**

The experimental data above suggest that there exist at least two degenerate states on the 1/3 and 2/3 plateaus at low temperatures. First we focus on the 1/3 plateau and discuss the 2/3 plateau later. We have pointed out in Ref. 38 that besides the state depicted at the top of Fig. 1a (denoted by $S_{1/3}$ hereinbelow; also see Fig. S5c), there exists another ice-rule state $S'_{1/3}$ (Fig. S5d, equivalent to the state depicted below $S_{1/3}$ in Fig. 1a) with the same net magnetization and the same refinement factors (Fig. S6). An exhaustive search for all $\sqrt{3} \times \sqrt{3}$ ice-rule states with the same net magnetization confirmed that $S_{1/3}$ and $S'_{1/3}$ are the only distinct states on the 1/3 plateau. Fig. 4a shows that the two states can be visualized as a 6-spin ring with opposite chiralities enclosed in a larger 3-spin triangle in a 2-in-1-out state. $S_{1/3}$ and $S'_{1/3}$ are distinct since they cannot be made equivalent by any operations in their magnetic space group Am'm2'. Moreover, they have the same energy based on the classical spin model in Ref. 38, which involves up to 4th neighbor exchange couplings and dipolar interactions among the Ho spins (Supplementary Note 6)[41].

To see whether $S_{1/3}$ and $S'_{1/3}$ can have different AHE, we constructed a toy model respecting the symmetries of HoAgGe and directly calculated the intrinsic contribution to the anomalous Hall conductivity. The toy model (Supplementary Note 4)[41] describes s-electrons hopping between nearest-neighboring Ho sites subject to local exchange fields parallel to the ordered Ho moments and to spin-orbit coupling determined by the inversion symmetry breaking with respect to the nearest-neighbor bond centers. Surprisingly, despite the complex lattice structure and magnetic order, the band structures for $S_{1/3}$ and $S'_{1/3}$ are completely identical (Fig. 4b). While this may be due to the model being oversimplified, the exact degeneracy of the Bloch band structures implies nontrivial symmetry relations between $S_{1/3}$ and $S'_{1/3}$, which will be discussed in detail below. What is more interesting, however, is that the two states generally have different Berry curvatures (summed over occupied bands) at the same positions in the Brillouin zone (Figs. 4b, c). Consequently, integrating the Berry curvature over the Brillouin zone gives different intrinsic anomalous Hall conductivities for $S_{1/3}$ and $S'_{1/3}$ (Fig. 4d).

**Origin of the AHE size difference**

To understand why the symmetry operation $\mathcal{X}$ that connects $S_{1/3}$ and $S'_{1/3}$ preserves the band structure but not the anomalous Hall effect, we first note that $\mathcal{X}$ can be represented by $R_b^\pi \mathcal{D}$, where $R_b^\pi$ is a usual $\pi$-rotation about the $b$ axis by $\pi$ on both orbital and spin degrees of freedom, and $\mathcal{D}$ is a special operation that reverses the distortion of the lattice but keeps the spin degrees of freedom untouched (Fig. 4a; also see Fig. 1a). In the paramagnetic state $\mathcal{D}$ is equivalent to a vertical mirror parallel to the $b$ axis, or a $\pi$-rotation about the $c$ axis, but cannot be represented by a standard magnetic space group operation when the magnetic order on the 1/3 plateau is taken into account. We note that distortion reversal has been suggested to be included as an additional "color symmetry" operation in crystalline space groups[39, 40]. In nonmagnetic crystals distortion reversal can usually be made equivalent to other space group operations such as mirror reflection, which transform physical observables in well-defined manner. The effect of $\mathcal{D}$ on physical quantities such as the Hamiltonian and the AHE needs to be analysed case-by-case due to its separate actions on orbital and spin degrees of freedom. In Supplementary

Note 5 we prove that $\mathcal{D}$ leaves all terms in our tight-binding model invariant except for changing the sign of the crystal momentum along $c$, thereby time-reversal-like. $\mathcal{X} = R_b^\pi \mathcal{D}$ thus keeps the band structure invariant additionally due to the $\mathcal{T}R_a^\pi$ symmetry in Am'm2'. We also prove that the classical spin model that successfully captured the various phases of HoAgGe gives identical energies for $S_{1/3}$ and $S'_{1/3}$ even with the long-range dipolar interaction considered. Therefore, the two states are degenerate in the same magnetic field range corresponding to the 1/3 plateau.

The reason why the *size* of the anomalous Hall conductivity is *not* invariant under $\mathcal{X}$ is more subtle. In the SI we show that the distortion reversal $\mathcal{D}$ critically changes the Berry connection, or the position operator's matrix elements between Bloch states, due to the shift of sublattice positions under $\mathcal{D}$, which changes the Berry curvature of occupied states and hence the anomalous Hall conductivity. Such a contribution is separate from the transformation of the Hamiltonian under $\mathcal{D}$, and is a manifestation of the fact that the Berry curvature encodes geometric information of the Bloch wavefunctions beyond the Hamiltonian itself. We note that the above mechanism will generally be relevant in structurally distorted magnetic systems in which the distortion reversal operation is not equivalent to a magnetic space group operation.

**Possible degenerate states on the 2/3 plateau**

We finally turn to the 2/3 plateau. Despite the same Am'm2' symmetry as the 1/3 plateau states, the 2/3 plateau only has one nondegenerate state $S_{2/3}$ (arrows in Fig. 4e; also see Fig. S5e). Alternatively, the resulting state by acting $\mathcal{X}$ on $S_{2/3}$ is equivalent to $S_{2/3}$ under a translation. A possible explanation is that the Ho spins are not exactly Ising[38], and below 4K the out-of-plane components of the Ho spins also become ordered[28]. The quantum spin Hamiltonian of HoAgGe involving the full low-energy CEF multiplet structure and the resulting low-temperature order will be reported in a future work. Here we only show that in the presence of an out-of-plane components order illustrated in Fig. 4e, the two states connected by $\mathcal{X}$ become inequivalent and again give different anomalous Hall conductivities (Fig. 4f), although the total energies and band structures are still the same (Supplementary Note 7). Awaiting future experimental and theoretical evidence, the particular finite-wave-vector out-of-plane order considered here[28] is reasonable due to the zero out-of-plane net magnetization down to 1.8K (see Fig. S2E of Ref. 38), suggesting dominant antiferromagnetic couplings between the out-of-plane spin components. A similar out-of-plane order for the 1/3 plateau can also be added to $S_{1/3}$ and $S'_{1/3}$ which does not qualitatively change the discussion above (Supplementary Note 7).

**Discussion and conclusion**

The existence of degenerate states alone does not explain why there is hysteresis across the whole 1/3 and 2/3 plateaus, since the degeneracy is not lifted by the external magnetic field as in the hysteresis of a ferromagnet. A possible explanation is that the barriers and kinetic pathways between the degenerate states and those on a neighboring plateau are not the same. Starting from a uniform state on a neighboring plateau and gradually changing field, transition to the degenerate state with a lower kinetic barrier will be dominant at low temperatures. The hysteresis at a given plateau will then appear if the transition from a higher-field plateau and that from a lower-field plateau do not favor the same state kinetically. For noncollinear magnetic order the kinetic pathways between nearly degenerate states are expected to be more complex than collinear magnets[24, 25]. Future studies done with high-resolution magnetic imaging techniques will hopefully reveal more details on the phase boundaries at the metamagnetic transitions in HoAgGe.

In conclusion, with the help of electrical transport measurements in the metallic kagome ice compound HoAgGe at low temperatures, we identified an emergent time-reversal-like degeneracy between metastable ice-rule states that have the same net magnetization but different anomalous Hall effect. The different sizes of the intrinsic anomalous Hall conductivity can be traced to the distortion-induced Berry connection of the Bloch electrons. We expect that similar phenomena should commonly appear in frustrated spin systems with nontrivial structural distortion and finite ordering wave vectors. The appearance of opposite toroidal moments on the two degenerate 1/3 plateau states also points to a mechanism for using the AHE to distinguish toroidal orders. Our study suggests a significant potential for transport measurements in distinguishing hidden symmetries in metallic frustrated spin systems.

**Acknowledgements**

The authors would like to thank Oleg Tchernyshyov, Masafumi Udagawa, Hao Deng, and Istvan Kezsmarki for helpful discussion. The work in Augsburg was supported by the German Research Foundation (DFG) through SPP1666 (project no. 220179758), TRR80 (project no. 107745057) and via the Sino-German Cooperation Group on Emergent Correlated Matter. K.Z. acknowledges the support by the Beijing Nova Program (Grant No. Z211100002121095). H.C. acknowledges the support by NSF CAREER grant DMR-1945023.

**Author contribution**

K.Z. and P.G. proposed the experiments; K.Z. synthesized single crystals and measured magnetic properties, specific heat, and electronic transport properties; Y.T conducted magnetic Grüneisen parameter measurements; H.C. proposed the theoretical analysis and provided model calculations; K.Z., H.C., and P.G. wrote the manuscript with input from all authors.

## Methods

### Single crystal growth and characterization

High quality single crystals of HoAgGe were grown using the Ag–Ge-rich self-flux method, with typical concentration $R_{0.06}(Ag_{0.75}Ge_{0.25})_{0.94}$ (R: Ho and Lu). Mixtures were placed in alumina crucibles and sealed in a quartz tube, heated to 1150°C, held there for 10h and cooled to 836°C within 76 hours, where the flux was decanted using a centrifuge.

All single crystals were characterized via x-ray diffraction with a Rigaku X'pert diffractometer using Cu $K\alpha_1$ radiation. Magnetization (M) measurements were performed using a vibrating sample magnetometer (Quantum Design MPMS). Electronic transport and specific heat data were obtained with Quantum Design PPMS.

### Demagnetization correction

A photo of the HoAgGe crystal for Hall effect measurements is included to Fig. S1. The $b$ axis is perpendicular to the plane, with thickness 0.5mm, and 1.4 mm length along the $c$ axis. Due to slightly different length of $a$ axis between the top and bottom surface, we adopt the average length of $a$ axis as 1.1 mm. The demagnetizing factor of this crystal is calculated by approximating it by a rectangular cuboid and using the corresponding formula in Ref. 42, which yields D=0.6. The demagnetization correction is $H_{int}= H_{ext}-$ D·M, where D is the demagnetization factor, and M the measured magnetization. The magnetization curve in panel A of Fig. S1 shows the raw data, while curve in panel B shows the one after correction for the internal demagnetizing field.

### Thermodynamic measurements

The magnetic specific heat $C_{mag}$ data in Fig. 4 and Fig. S4 were obtained by subtracting the contributions from Ho nuclei, lattice vibrations, and itinerant electrons, as discussed in Ref. 38. The ordering temperature around 7K in Fig. S4(a) is determined by the mid-point of the transition signature

The magnetic Grüneisen parameter or adiabatic magnetocaloric effect was measured using an alternating field technique adapted to a dilution refrigerator, see Ref. 43. Note, that $\Gamma_{mag} \equiv \frac{1}{T}\left(\frac{dT}{d\mu_0 H}\right)_S = -\frac{1}{\mu_0 C}\frac{dS}{dH}$ which can be approximated by $-\frac{1}{\mu_0 C_{mag}}\frac{dS_{mag}}{dH}$ since the non-magnetic specific heat is much smaller than the magnetic one even at 12 K, cf. Fig. 3A of ref. 38. This approximation allows to estimate the field-dependence of magnetic entropy by integration.

**Magnetic structure refinement**

Elastic single crystal neutron scattering details have been clearly explained in the supplementary materials of Ref. 38. The magnetic structures are determined using Jana 2006 software[44].

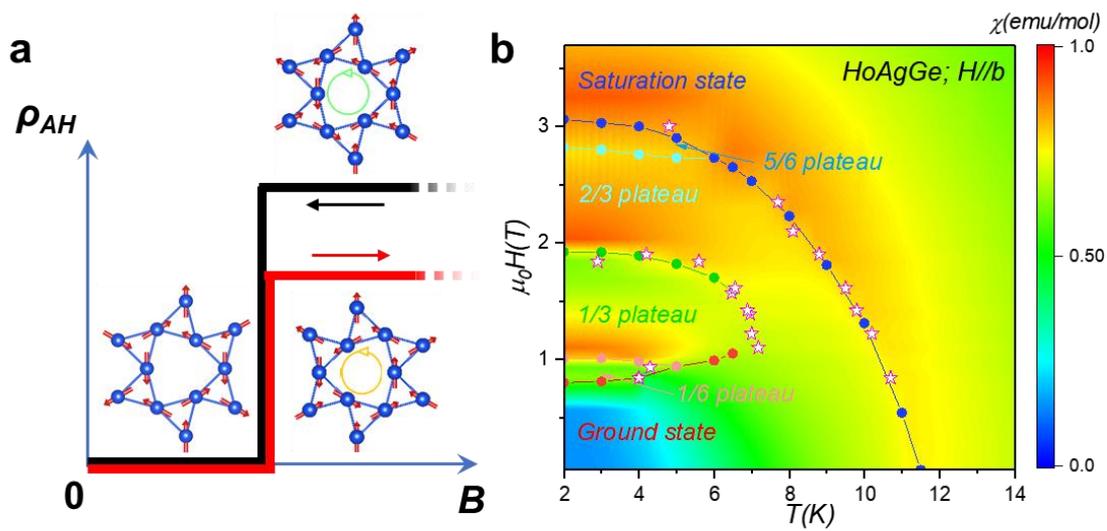

**Fig. 1| Time-reversal-like degeneracies in field-induced plateau phases of HoAgGe.**
**a,** illustration of the finite-field hysteresis of the anomalous Hall resistivity *size* due to emergent time-reversal-like degenerate states. The two states on the finite-field plateau are respectively $S_{1/3}$ (top) and $S'_{1/3}$ (bottom right, transformed by $R_b^\pi$ for better illustration) defined in the text. Also depicted is the kagome ice ground state which has zero AHE (bottom left). **b,** The H-T phase diagram of HoAgGe under *H//b* as derived from *M(H)* (filled circles) and $C_{mag}$ and $\Gamma_H$ (empty stars) measurements. The color coding represents the magnetic susceptibility (see text).

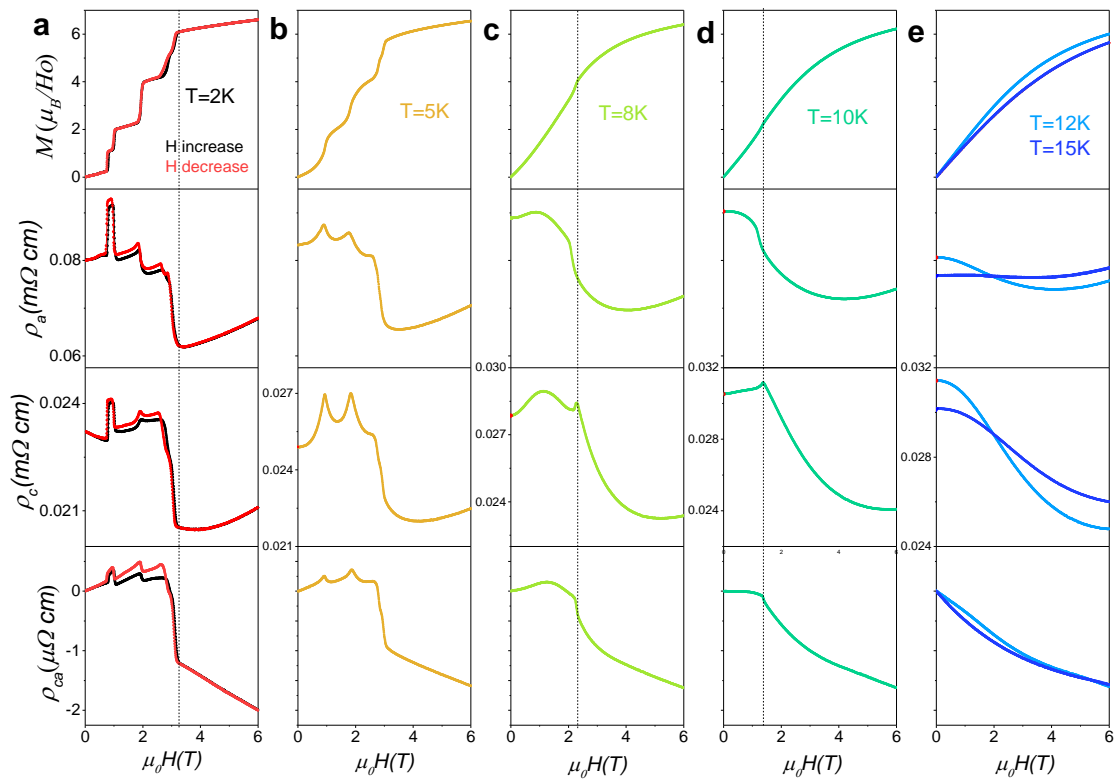

**FIG. 2| Magnetic and transport properties of HoAgGe.** For H//*b*, the Magnetization (M), *a*- and *c*-axis magnetoresistance $\rho_a$ and $\rho_c$, as well as the Hall resistivity $\rho_{ca}$ (current along *c*, voltage along *a*) are displayed at 2K (**a**), 5K (**b**), 8K (**c**), 10K (**d**), 12K and 15K (**e**), respectively. Dashed lines indicate the saturation fields.

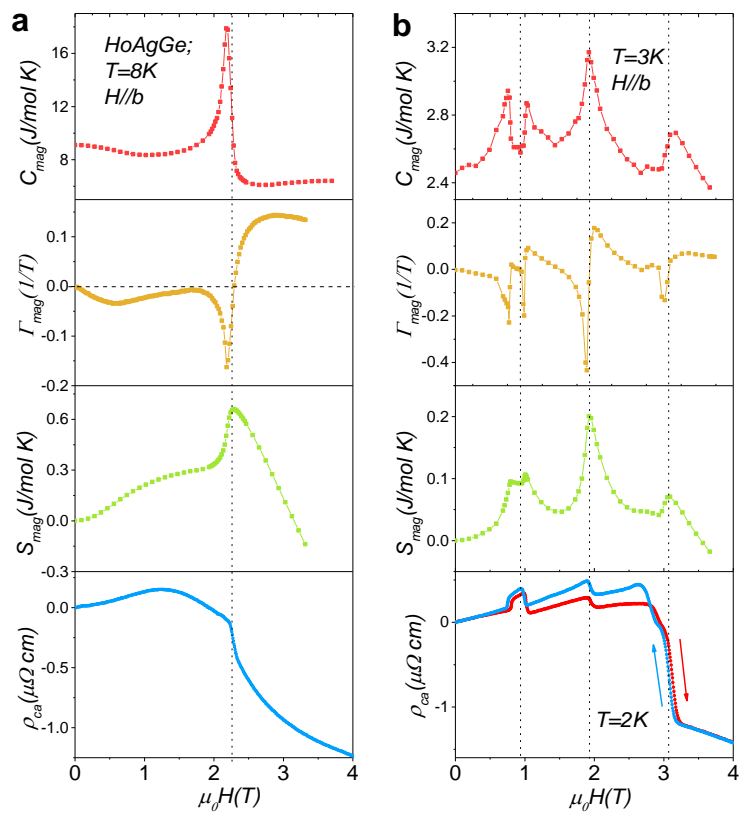

**Fig. 3| Thermodynamic measurements of HoAgGe in comparison with the AHE.**
Field dependence of magnetic specific heat $C_{mag}$, magnetic Grüneisen parameter $\Gamma_H$, magnetic entropy $S_{mag}$, and AHE of HoAgGe at 8K (a), and 3K (2K for AHE) (b), under H//b.

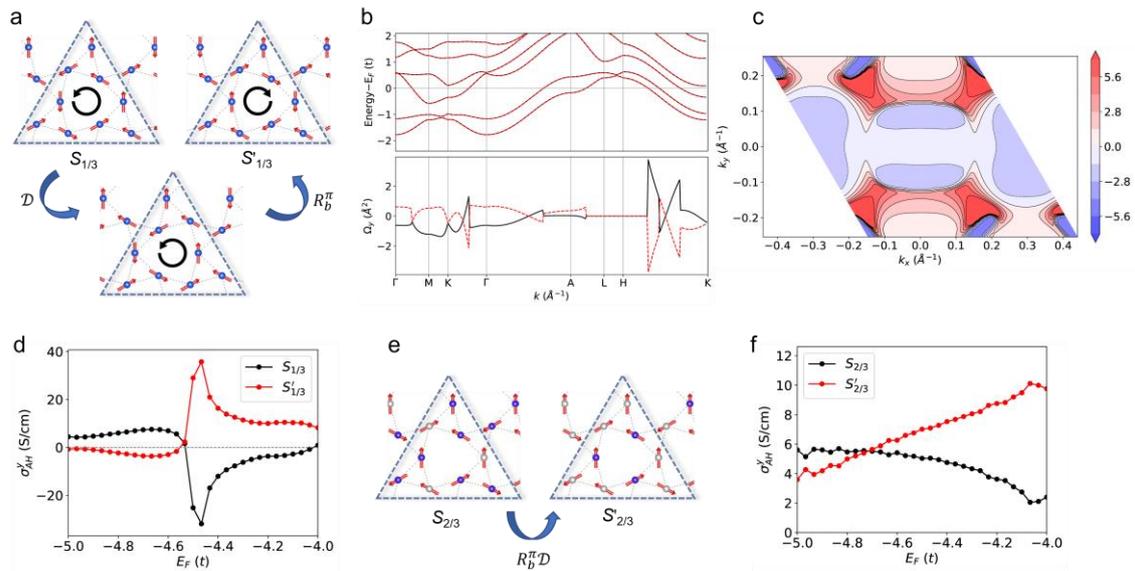

**Fig. 4| Model calculations of the anomalous Hall effect on the 1/3 and 2/3 plateaus.**
**a,** Two degenerate states $S_{1/3}$ and $S'_{1/3}$ on the 1/3 plateau, connected by $\mathcal{X} = R_b^\pi \mathcal{D}$. **b,**
Comparison between the band structures (top panel) and Berry curvatures summed over
occupied bands (bottom panel) of $S_{1/3}$ (black solid lines) and $S'_{1/3}$ (red dashed lines).
The parameter values are $t_v = -0.4$, $\lambda_h = 0.3$, $\lambda_v = 0.2$, $J = -2.2$, $\mu = -4.4$, where
$t_v, \lambda_h, \lambda_v, J, \mu$ are nearest-neighbor out-of-plane hopping, in-plane spin-orbit coupling,
out-of-plane spin-orbit coupling, local exchange coupling, and chemical potential,
respectively. The nearest-neighbor in-plane hopping is fixed to $-1$. **c,** Difference between
the Berry curvatures of $S_{1/3}$ and $S'_{1/3}$ in the $k_z = 0$ plane. **d,** Intrinsic anomalous Hall
conductivities of $S_{1/3}$ and $S'_{1/3}$ versus Fermi energy (or $\mu$). **e,** Possible degenerate states
on the 2/3 plateau, $S_{2/3}$ and $S'_{2/3}$, with out-of-plane spin components ordering. Grey and
blue spheres correspond to spins pointing into and out of the paper, respectively. **f,**
Intrinsic anomalous Hall conductivities of $S_{2/3}$ and $S'_{2/3}$ versus Fermi energy.

# Supplementary Information for "Time-reversal-like degeneracies distinguished by the anomalous Hall effect in a metallic kagome ice compound"

K. Zhao[1,2] , Y. Tokiwa[1], H. Chen[3,4] , and P. Gegenwart[1]

*1. Experimentalphysik VI, Center for Electronic Correlations and Magnetism,*

*University of Augsburg, 86159 Augsburg, Germany*

*2. School of Physics, Beihang University, Beijing 100191, China*

*3. Department of Physics, Colorado State University, Fort Collins, CO 80523-1875, USA*

*4. School of Advanced Materials Discovery, Fort Collins, CO 80523-1875, USA*

**Table of Content**

**Supplementary figures in Method part**

**Supplementary Note 1: Neutron refinement of degenerate states at 1/3 plateau, and test about 2/3 plateau case**

**Supplementary Note 2: Electronic transport properties under H//b and H//a of HoAgGe**

**Supplementary Note 3: Time-reversal-like degeneracies**

**Supplementary Note 4: Toy model of HoAgGe**

**Supplementary Note 5: Transformation of the Hamiltonian and the AHE under distortion reversal**

**Supplementary Note 6: Transformation of the classical spin model under distortion reversal**

**Supplementary Note 7: Degenerate states with nonzero out-of-plane spin components**

**Supplementary figures in Method part**

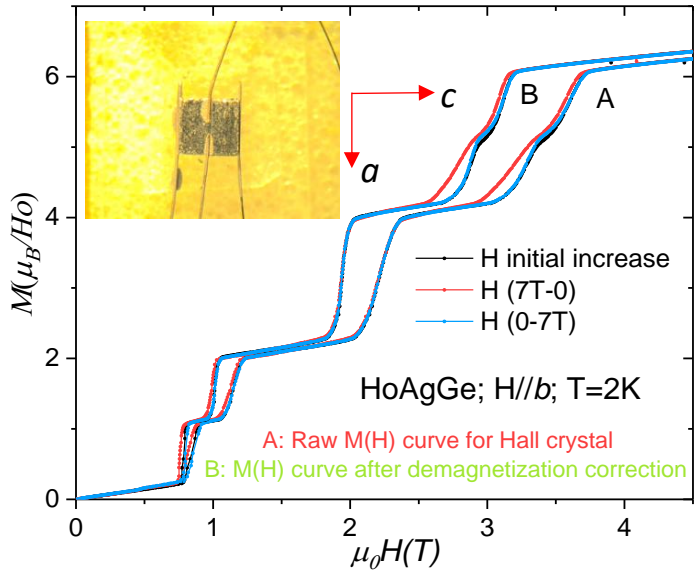

Fig. S1. Raw and demagnetization corrected isothermal in-plane (H//b) magnetization of HoAgGe for Hall measurements at 2K, with photo of the crystal (see text).

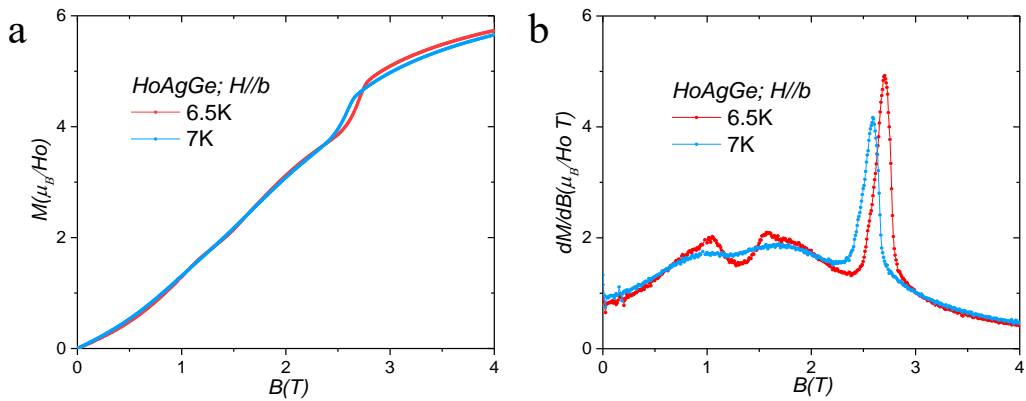

Fig. S2. Isothermal in-plane (**H**//*b*) magnetization (a) and corresponding differential susceptibility *dM/dH* (b) for HoAgGe at 7K and 6.5K, with three peaks for the 6.5K data, indicated in Fig. 2a.

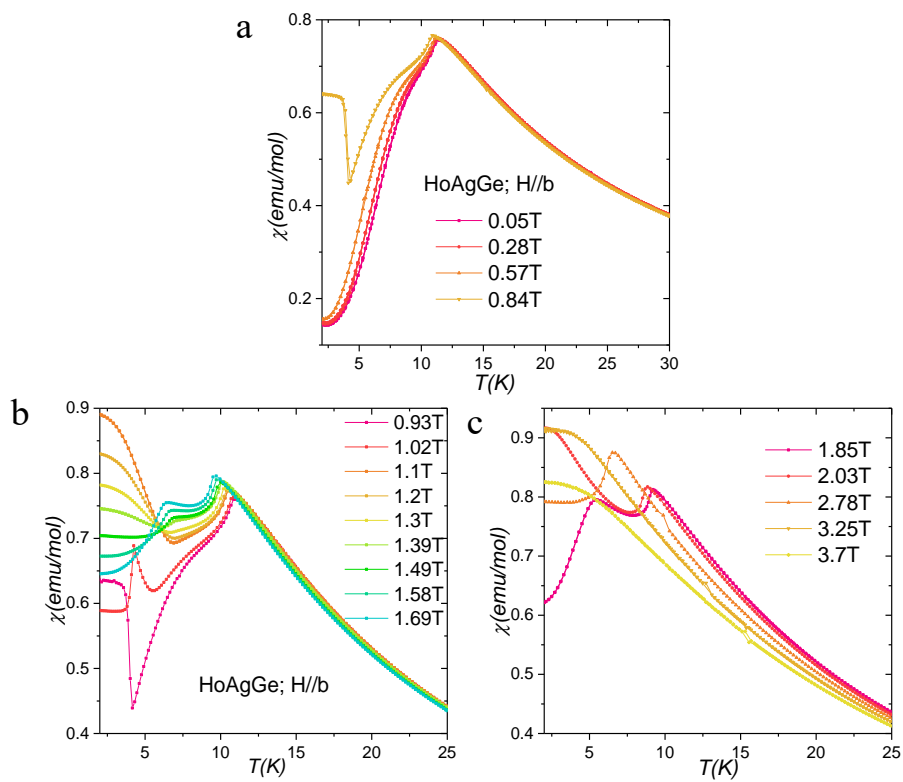

Fig. S3. Low-temperature susceptibility χ(T) of HoAgGe for **H**//*b*, under various magnetic fields.

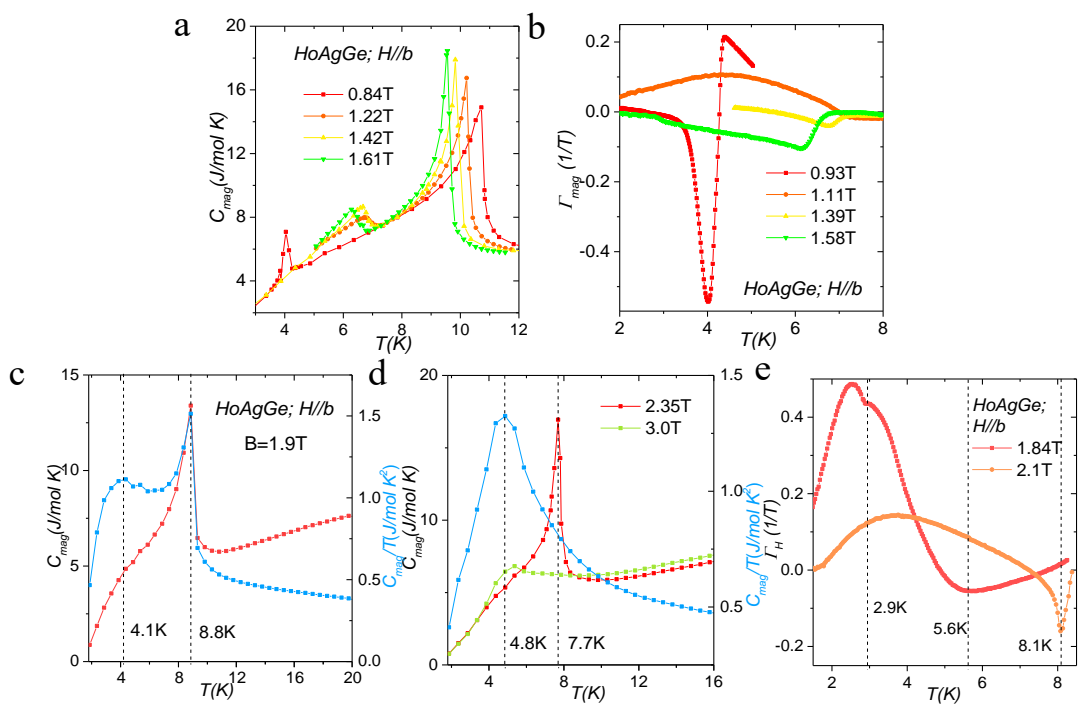

Fig. S4. **a** and **b** display the temperature dependence of the magnetic specific heat $C_{\text{mag}}$ and magnetic Grüneisen parameter $\Gamma_H$ under **H**//*b* at various magnetic fields, respectively. **c,** Magnetic specific heat $C_{\text{mag}}$ and $C_{\text{mag}}$/T data of HoAgGe at 1.9T under **H**//*b* below 20K. **d,** Magnetic specific heat $C_{\text{mag}}$ of HoAgGe at 2.35T and 3T under **H**//*b* below 16K, together with $C_{\text{mag}}$/T data at 3T as right panel. **e,** Magnetic Grüneisen parameter $\Gamma_H$ of HoAgGe at 1.84T and 2.1T under **H**//*b* below 8.5K, with dashed lines marking the transition temperatures.

## Supplementary Note 1: Neutron scattering analysis of the 1/3 and 2/3 plateau states

Fig. S5 shows the ground state of the kagome spin ice ground state (a) and saturation state (b) under H//b of HoAgGe below 4K. Under H//b, the magnetic field breaks the three-fold rotational symmetry and reduces the P-6′m2′ symmetry of the ground state (Fig. S5(a) with single domain) to Am′m2′ in the 1/3 plateau (Fig. S5(c-d) with three domains), with the nine Ho moments in the $\sqrt{3} \times \sqrt{3}$ unit cell forming six non-equivalent groups.

Fig. S5(c) shows the magnetic pattern for Ho4 flipping ($S_{1/3}$ state) under 1.5T at 4K, namely on the one-third plateau. In fact, either Ho4 or Ho6 flipping ($S'_{1/3}$ state in Fig. S5(d)) would give the same magnetic refinement factors, with *R=13.85%* and

*wR=17.31%,* under the constraint of three equivalent domains in Fig. S6(a). When further relaxing this constraint, the refinement factor would be reduced to the values of *R=8.61%* and *wR=10.85%,* as shown in Table. 1 of Ref. 1.

Compared to Fig. 1A of Ref. 1, our recent magnetic susceptibility $\chi(T)$ results down to 1.8 K with denser data points in Fig. S7 clearly show that, in addition to the sharp peak at 11.8 K and broad peaks at ~7 K, another transition at 2.5 K, for both **H**//*b* and **H**//*c* under 500 Oe occurs. Compared with **H**//*b* axis, $\chi(T)$ exhibits more clear increase under **H**//*c* axis, indicating that the out-of-plane ($M_c$) components of the Ho spins also become ordered around 2.5K.

As shown in Fig. S8A of Ref. 1, the scattering vector Q-scan plots along (h, -2h, 1) (equivalent to (h, h, 1)) direction, show only the magnetic vector (1/3, 1/3, 0) at 1.8K with H//*b* under H<4T. Thus, the $M_c$ components should become ordered with the same magnetic vector (1/3, 1/3, 0) at zero field and under H<4T. Under H = 4T, the magnetic unit cell of $M_c$ components should also be identical to the structural one, due to the disappearance of magnetic vector (1/3, 1/3, 0) in the saturation state.

With both Ho4 and Ho6 flipped, the magnetic refinement factors would be *R=12.37%* and *wR=15.47%,* with three equivalent domains. If further relaxing this constraint, the refinement factor would be reduced to the values of *R=5.90%* and *wR=7.19%* as shown in Table. 1 of Ref. 1. Fig. S5(e) shows the proposed magnetic pattern for the $S_{2/3}$ state under 2.5T at 1.8K, namely on the 2/3 plateau. Due to the $M_c$ component ordering, the magnetic space group (MSG) changes from Am′m2′ into C2′, with Ho3-Ho6 pointing out of the paper, and Ho1-Ho2 pointing into the paper, respectively. With equivalent domains, the magnetic refinement factors would be *R=12.54%* and *wR=16.77%* (Fig. S6(b)), rather close to the case without the $M_c$ component ordering.

The $M_{ab}$ and $M_c$ components are determined to be 7.7(1) $\mu_B$, and 1.5(6) $\mu_B$, respectively. For a given magnetic structure, the ordered moment *M* is proportional to the magnetic structure factor *F*. Since the magnetic peak integrated intensity $I = F^2$, *I* is proportional to $M^2$, which previously was $M_{ab}^2$ if only considering the *ab* plane components and now is $M_{ab}^2 + M_c^2$. Thus, it is reasonable that the neutron refinement only accounting for the $M_{ab}^2$ component still has reasonable refinement factors as listed in Table 1 of Ref. 1. As discussed in the main text, such a small $M_c$ component is enough to generate two degenerate states with different AHE size in the 2/3 plateau region.

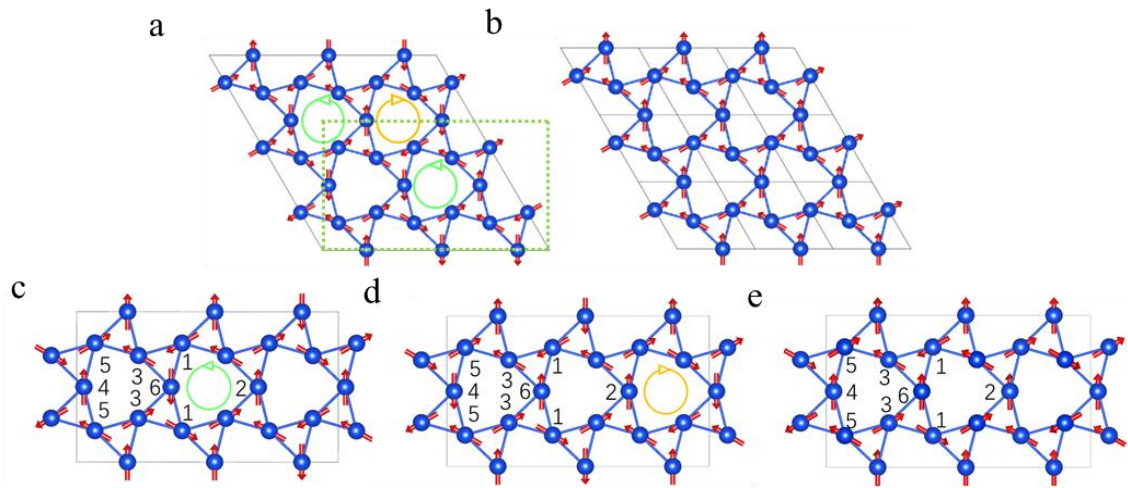

Fig. S5. **a,** Magnetic structure of HoAgGe at 4 K, namely the ground state of kagome spin ice in the $3 \times 3$ magnetic cell, according to Ref. 1. **b,** Magnetic structure of HoAgGe at H = 4 T and T = 1.8 K, i.e., the saturation state. Magnetic structure on the 1/3 plateau (H = 1.5 T and T = 4 K), with Ho4 flipping (**c**) and Ho6 flipping (**d**) as two degenerate states. **e,** Proposed magnetic structure at 2/3 plateau (H = 2.5 T and T = 1.8 K). The refinement was done in the $3 \times \sqrt{3}$ light-green rectangle, as shown in (**a**). The six inequivalent Ho sites are labelled by numbers 1 to 6 for simplicity in (**c-e**).

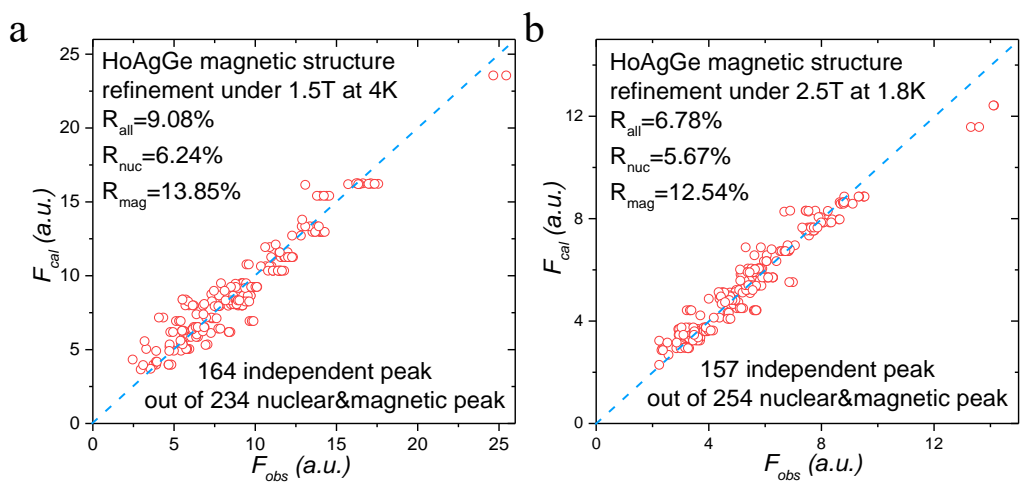

Fig. S6: Plots of calculated vs. experimental magnetic structure factors for the refined magnetic structures at 4K under 1.5T (**a**) and 1.8K under 2.5T (**b**) of HoAgGe.

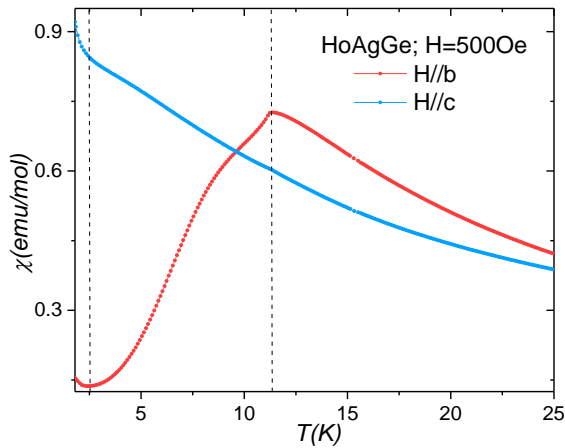

Fig. S7: Magnetic susceptibility of HoAgGe under H along *c* axis and *b* axis from 25 K to 1.8 K.

## Supplementary Note 2: Electronic transport properties for H//b and H//a in HoAgGe

Two magnetic transitions also show up in the resistivity curve, marked as $T_1$ (7K) and $T_2$ (11.5K) in the plot of the temperature derivative of $\rho(T)$ in Fig. S8. Experimentally we do not see evidence of strong anisotropy of the Fermi surface, as our transport measurements show comparable resistivity for currents in the ab plane and along the c axis, consistent with the neutron result of 3D Ising spins. Consistent with $M_c$ components ordering, there seems to be an anomaly below 4K, the origin of which is under current investigation.

We adopt a two-band-model fitting to estimate the carrier density in the paramagnetic state at 50K in Fig. S9a. Two electron bands are band 1: $1.0 \times 10^{19}$/cm$^3$ with mobility 1624 cm$^2$/V·s and band 2: $1.2 \times 10^{22}$/cm$^3$ with mobility 13 cm$^2$/V·s.

To determine the possibility of unconventional AHE that has been observed in various non-collinear magnets, we directly plot the Hall resistivity $\rho_{ca}$ versus M in Fig. S9b. If dominated by the conventional AHE, the Hall resistivity $\rho_{ca}$ should vary linearly with M, which is clearly observed at 15K and 12K. The saturation state also shows a linear dependence at 2 K, while it seriously deviates from linear behavior in the low field region, indicating another dominant contribution to the AHE, similar to the case of $Pr_2Ir_2O_7$[2], i.e., the third term in the naive decomposition $\rho_{xy} = R_0 H + R s \rho_{xx}^2 M + \rho_{xy}^{spin}$.

In a skyrmion lattice, the topological Hall effect (THE) is induced by the topological spin texture; and the system stays in the same magnetic state before and after the skyrimon phase, which enables the quantitative analysis of THE signal [3]. However, such a method is not directly applicable for the HoAgGe system, since the metamagnetic phase transitions would also strongly influence the band structure at the Fermi surface.

Furthermore, we analyzed the transport signal in terms of the conductivity tensor in Fig. S10a. The Hall conductivity is obtained using $\sigma_{ac} = \rho_{ca}/\rho_a \rho_c$ (valid in the case of $\rho_a$ $(\rho_c) \gg \rho_{ca}$), and shows hysteresis as well. Note that such a feature only appears below 4K in Fig. S10b, where kagome spin ice is fully ordered into the ground state.

Fig.S11a summarizes the magnetization (M) at 2K, and Hall resistivity $\rho_{cb}$ with current along c, voltage along b of HoAgGe at 15K and 2K, and H//a. Interestingly, the Hall resistivity $\rho_{cb}$ has small hysteresis for the 1/3 plateaus, in contrast to the indiscernible hysteresis in the magnetization curve at 2K. Fig. S11b directly plots the Hall resistance $\rho_{ca}$ versus M at 2K under H//a.

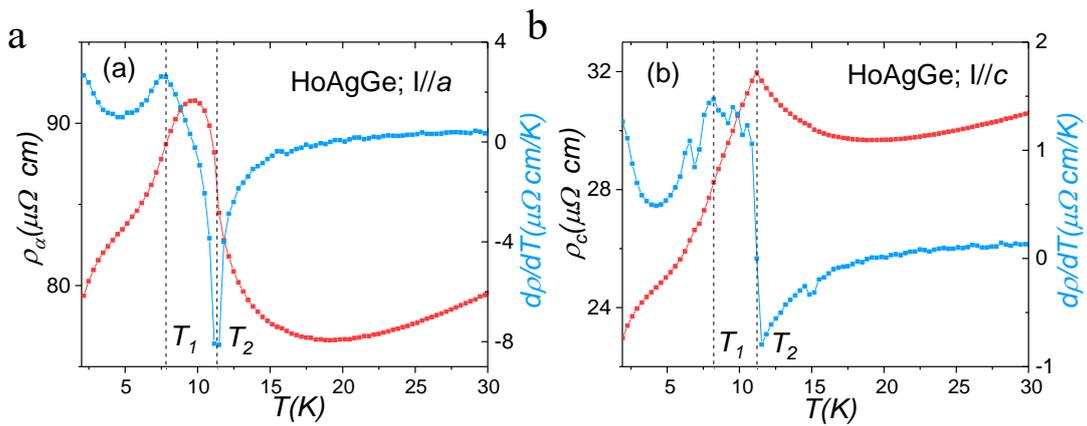

FIG. S8. Resistivity curve $\rho$(T) of HoAgGe from 30K to 2K, with current along (**a**) *a* axis, and (**b**) along *c* axis, together with corresponding temperature derivative of $\rho$(T) at the right axes.

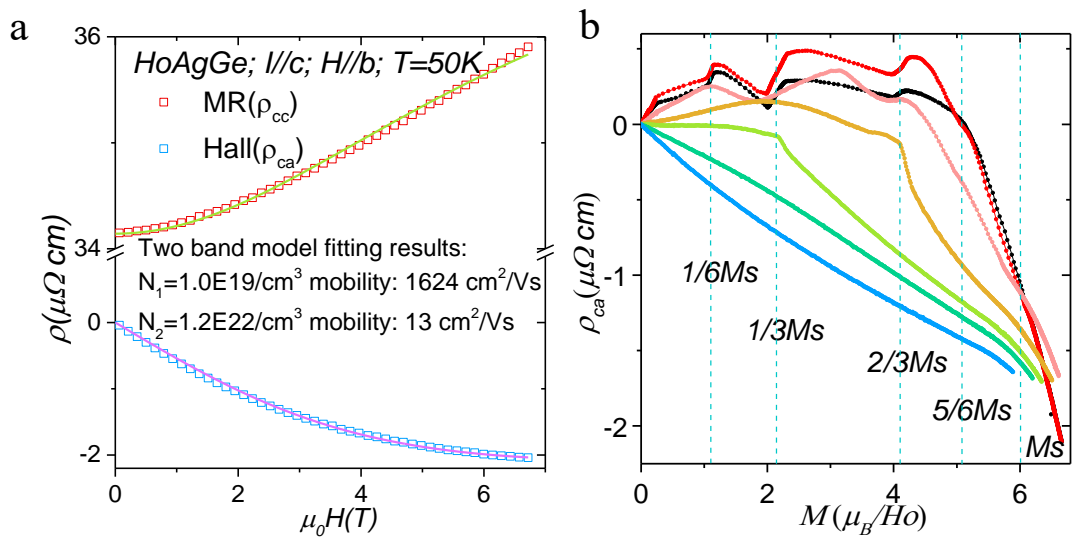

FIG. S9. **a**, Hall resistivity $\rho_{ca}$ and magnetoresistance $\rho_{cc}$ at 50K, with two band fitting (see text). **b**, Hall resistivity ca vs. magnetization M of HoAgGe are displayed at 2K, 5K, 8K, 10K, 12K and 15K, respectively.

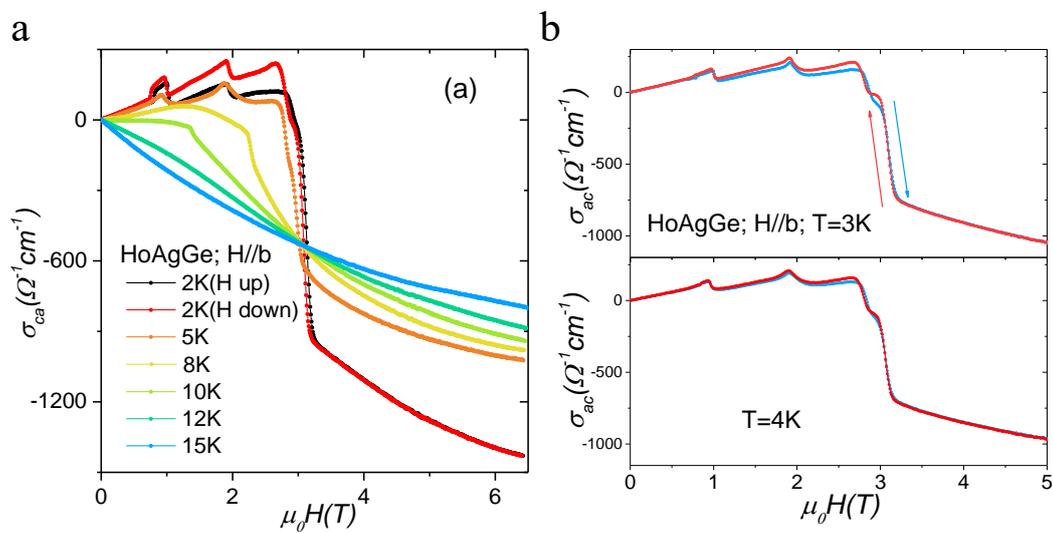

Fig.S10. **a** Field dependence of Hall conductivity $\sigma_{ca}=\rho_{ca}/\rho_c\rho_a$, of HoAgGe under H//b are displayed at 2K, 5K, 8K, 10K, 12K and 15K, respectively. **b** Hall conductivity at 3K (top panel) and 4K (bottom panel).

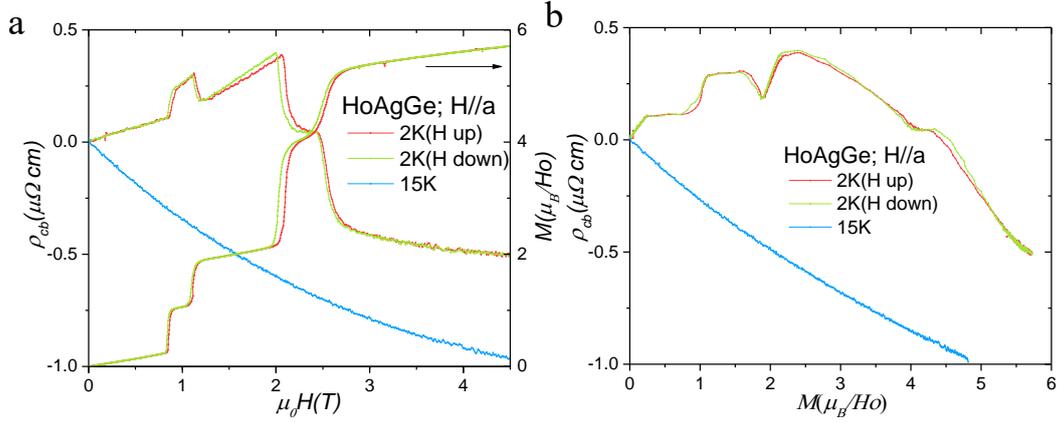

FIG. S11. **a** Hall resistivity $\rho_{cb}$ (current along c, voltage along b) of HoAgGe at 2K and 15K, with magnetization M under H//a at 2K as right panel. **b** Hall resistivity $\rho_{cb}$ vs. magnetization M of HoAgGe are displayed at 2K and 15K.

## Supplementary Note 3: Time-reversal-like degeneracies

We focus on the effect of time reversal $\mathcal{T}$ on the eigenenergies of a mean-field Hamiltonian. The total energy for a mean-field Hamiltonian is

$$E = \mathrm{Tr}[\rho H], \tag{1}$$

where $\rho$ is the density matrix. It is then clear that, by using the property of trace,

$$E = \mathrm{Tr}[\rho(H)H] = \mathrm{Tr}[\rho(\mathcal{T}H\mathcal{T}^{-1})\mathcal{T}H\mathcal{T}^{-1}] \tag{2}$$

In other words, two systems with the Hamiltonian $H$ and $\mathcal{T}H\mathcal{T}^{-1}$ have the same total energy, regardless of the TR properties of $H$.

One can also establish the relationship between the band structures of $H$ and $\mathcal{T}H\mathcal{T}^{-1}$. Suppose $|\psi_{n\mathbf{k}}\rangle$ is a Bloch eigenstate of $H$ with band energy $\epsilon_{n\mathbf{k}}$, we have

$$\mathcal{T}H\mathcal{T}^{-1}\mathcal{T}|\psi_{n\mathbf{k}}\rangle = \epsilon_{n\mathbf{k}}\mathcal{T}|\psi_{n\mathbf{k}}\rangle. \tag{3}$$

Namely, $\mathcal{T}|\psi_{n\mathbf{k}}\rangle$ is an eigenstate of $\mathcal{T}H\mathcal{T}^{-1}$, with the same eigen-energy $\epsilon_{n\mathbf{k}}$. One may generally write $\mathcal{T}|\psi_{n\mathbf{k}}\rangle$ as $|\phi_{m\mathbf{k}'}\rangle$. The labeling of the bands is arbitrary, but $\mathbf{k}$ and $\mathbf{k}'$ are related to the eigenvalues of the discrete translation operator and must have a definite relationship. Considering $T_{\mathbf{R}}$ which stands for translation by an arbitrary lattice vector $\mathbf{R}$, we have

$$T_{\mathbf{R}}|\psi_{n\mathbf{k}}\rangle = e^{i\mathbf{k}\cdot\mathbf{R}}|\psi_{n\mathbf{k}}\rangle. \tag{4}$$

Applying $\mathcal{T}$ on both sides of the above equation, we have

$$\mathcal{T}T_{\mathbf{R}}\mathcal{T}^{-1}\mathcal{T}|\psi_{n\mathbf{k}}\rangle = T_{\mathbf{R}}\mathcal{T}|\psi_{n\mathbf{k}}\rangle = \mathcal{T}e^{i\mathbf{k}\cdot\mathbf{R}}|\psi_{n\mathbf{k}}\rangle = e^{-i\mathbf{k}\cdot\mathbf{R}}\mathcal{T}|\psi_{n\mathbf{k}}\rangle, \tag{5}$$

or equivalently

$$T_{\mathbf{R}}|\phi_{m\mathbf{k}'}\rangle = e^{-i\mathbf{k}\cdot\mathbf{R}}|\phi_{m\mathbf{k}'}\rangle. \tag{6}$$

Therefore, $\mathbf{k}' = -\mathbf{k}$. So the band structure of $\mathcal{T}H\mathcal{T}^{-1}$ is the same as that of $\mathcal{T}H\mathcal{T}^{-1}$ up to a transformation of $\mathbf{k} \to -\mathbf{k}$.

Now suppose we have a unitary or anti-unitary operator $\mathcal{X}$ whose form is dependent on the Hamiltonian (i.e., not universal), and we consider a particular form of the Hamiltonian

$$H = H_0 + H_M \tag{7}$$

where $H_0$ is the nonmagnetic part of the Hamiltonian and $H_M$ explicitly depends on the magnetic order parameter, which can be a general magnetization vector field $\mathbf{M}$. Due to the property of the trace we can still conclude that $H$ and $H' \equiv \mathcal{X}H\mathcal{X}^{-1}$ always have the same total energy. But a system described by $H'$ is in general very different from the system described by $H$. For example, if $\mathcal{X}$ involves interchanging certain atomic sites or distorting the lattice, we in general do not have $H_0' = H_0$.

On the other hand, however, if one can find a $\mathcal{X}$ so that (1) $H_0' = H_0$, and (2) $H_M' = H_M(\mathbf{M}')$, i.e., it only changes the magnetization vector field $\mathbf{M}$ to $\mathbf{M}'$, we will have a rather interesting situation that a degeneracy is guaranteed for two very different magnetic orders not related by time reversal. This degeneracy extends to the band structure as well, since

$$H'(\mathcal{X}|\psi_{n\mathbf{k}}\rangle) = \epsilon_{n\mathbf{k}}(\mathcal{X}|\psi_{n\mathbf{k}}\rangle) \equiv \epsilon_{n\mathbf{k}}|\phi_{n\mathbf{k}'}\rangle, \tag{8}$$

i.e. the Hamiltonian $H'$ has an eigenstate $|\phi_{n\mathbf{k}'}\rangle$ with the same band energy $\epsilon_{n\mathbf{k}}$. But in this case the relationship between $\mathbf{k}'$ and $\mathbf{k}$ may be more complicated depending on how $T_{\mathbf{R}}$ transforms under $\mathcal{X}$.

Another nontrivial consequence is that the two states may have very different physical properties. Consider an arbitrary observable $O$. We have

$$\langle O(H)\rangle = \mathrm{Tr}[\rho(H)O] = \mathrm{Tr}[\rho(H')O'] \neq \langle O(H')\rangle, \tag{9}$$

as long as $O' \neq O$. For vector or tensor objects that have well-defined transformation properties under space group operations, $O'$ and $O$ may not have as simple relations since $\mathcal{X}$ is not a space-group operation. The discussion can be extended to response functions as well. Thus it is possible that $H$ and $H'$ have different *sizes* of the anomalous Hall effect, spin magnetization, and orbital magnetization, etc., in contrast to the case of $\mathcal{T}$.

## Supplementary Note 4: Toy model of HoAgGe

In this section we give a toy model motivated by the plateau states of HoAgGe under a magnetic field along the $b$ axis. The model describes $s$-electrons hopping on the lattice formed by the Ho ions in HoAgGe:

$$H = \sum_{ij,\alpha} t_{ij} c_{i\alpha}^\dagger c_{j\alpha} + i \sum_{ij,\alpha\beta} \lambda_{ij} (\hat{e}_{ij} \times \hat{r}_{ij}) \cdot \boldsymbol{\sigma}_{\alpha\beta} c_{i\alpha}^\dagger c_{j\beta} + J \sum_{i,\alpha\beta} \hat{n}_i \cdot \boldsymbol{\sigma}_{\alpha\beta} c_{i\alpha}^\dagger c_{i\beta} \quad (10)$$

$$\equiv H_t + H_{so} + H_{ex}$$

where the three terms respectively correspond to spin-independent hopping, spin-orbit coupling, and exchange coupling with on-site Zeeman fields. In the above equation $i, j$ label lattice sites, $\alpha, \beta$ label spin, i is the imaginary unit. $t_{ij}$ is the spin-independent hopping amplitude between sites $i, j$, $\lambda_{ij}$ is the amplitude of spin-orbit coupling. $\hat{e}_{ij}$ is a vector standing for the electric field or electric dipole direction at the center of the $ij$ bond (see below), which is a simple way to introduce symmetry-allowed spin-orbit coupling for $s$-electrons. $\hat{r}_{ij}$ is the direction unit vector along the $ij$ bond. $J$ is the strength of the exchange coupling between the $s$-electrons and the local Zeeman field along $\hat{n}_i$ on site $i$. $\hat{n}_i$ will have the same pattern as that of the Ho local moments in a given magnetic configuration.

$\hat{e}_{ij}$ are determined in the following way. We first calculate the electric dipole moment with respect to the center of the $ij$ bond:

$$\mathbf{P}_{ij} = \sum_k Q_k [\mathbf{r}_k - (\mathbf{r}_i + \mathbf{r}_j)/2] \quad (11)$$

where $k$ runs through the nearest neighbors of sites $i, j$ including Ho, Ag, and Ge atoms. We assign $Q_{Ho} = +3$, $Q_{Ag} = +1$, and $Q_{Ge} = -4$. Then $\hat{e}_{ij} = \mathbf{P}_{ij}/|\mathbf{P}_{ij}|$. The simplified lattice structure for the tight-binding model is shown in Fig. S12. In Fig. S12 the directions and lengths of the arrows represent that of $\mathbf{P}_{ij}$. It is evident that the arrows have the same symmetry as the original HoAgGe lattice.

The results shown in Fig. 4 of the main text were obtained by using nearest-neighbor (either in-plane or out-of-plane) bonds in $H_t$ and $H_{so}$ only. To account for the different sizes of $\mathbf{P}_{ij}$ for in-plane and out-of-plane nearest neighbor bonds, we scale $\hat{e}_{ij}$ further by $|\mathbf{P}_{ij}|$ of the corresponding bond. Namely, the $\hat{e}_{ij}$ for out-of-plane nearest neighbor bonds are unit vectors, while that for in-plane ones have a length of 0.4152.

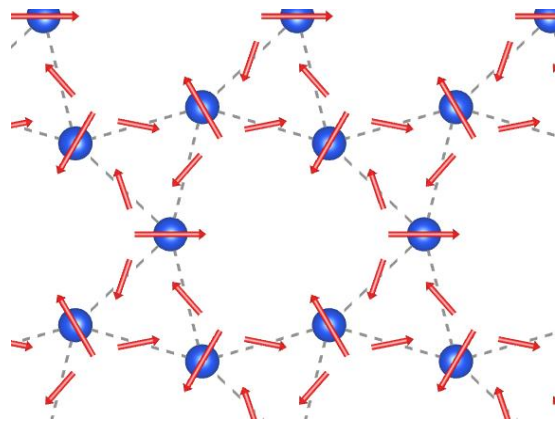

**Fig. S12| Lattice structure of the tight-binding model.** The arrows represent the directions and sizes of the electric dipole moments defined in Eq. (11).

## Supplementary Note 5: Transformation of the Hamiltonian and the AHE under distortion reversal

To understand how a structural transformation is reflected in a general tight-binding Hamiltonian, we consider a complete set of orthonormal Wannier functions $\{|a\rangle\}$. The tight-binding representation of an arbitrary non-interacting Hamiltonian is

$$H = \sum_{a,b} t_{ab}|a\rangle\langle b| \tag{12}$$

where $t_{ab} = \langle a|H|b\rangle$. For a standard $O(3)$ transformation, e.g. inversion $\mathcal{I}$, the Hamiltonian of the transformed system, independent of basis, is $\tilde{H} = \mathcal{I}H\mathcal{I}^{-1}$. One can write $\tilde{H}$ in the original Wannier basis of $H$, but the resulting hopping parameters will not have simple relations to $t_{ab}$. Instead, one can define a new Wannier basis $\{|\tilde{a}\rangle\} = \{\mathcal{I}|a\rangle\}$, so that

$$\tilde{H} = \sum_{ab} |\tilde{a}\rangle\langle a|\mathcal{I}^{-1}\mathcal{I}H\mathcal{I}^{-1}\mathcal{I}|b\rangle\langle \tilde{b}| = \sum_{ab} t_{ab}|\tilde{a}\rangle\langle \tilde{b}|. \tag{13}$$

Namely, the matrix elements of the Hamiltonian are unchanged as long as one regards the basis as that transformed by $\mathcal{I}$.

However, since the distortion reversal is not a space group operation, the transformed Hamiltonian, denoted as $H'$, in general does not have a definite relationship with $H$ without specifying the representation. If we use a Wannier basis based on the atomic sites of the distortion-reversed lattice, denoted by $\{|a'\rangle\}$, we can write

$$H' = \sum_{ab} |a'\rangle\langle a'|H'|b'\rangle\langle b'| \equiv \sum_{ab} t'_{ab}|a'\rangle\langle b'|. \tag{14}$$

Although the relationship between $H'$ and $H$ is unknown, one can use the same rules that lead to $t_{ab}$ to obtain the matrix elements $\langle a'|H'|b'\rangle$, thereby establishing an effective transformation between $H'$ and $H$ in the respective Wannier bases.

To this end we first consider how an arbitrary bond is modified by the distortion reversal in the present model. If the three sublattices of the original nonmagnetic unit cell are located at $\boldsymbol{\tau}_{1,2,3}$, the lattice vectors of the nonmagnetic lattice are

$$\mathbf{a}_1 = a\hat{x}, \ \ \mathbf{a}_2 = -\frac{1}{2}a\hat{x} + \frac{\sqrt{3}}{2}a\hat{y}. \tag{15}$$

Positions of $1', 2', 3'$ can be obtained as

$$\boldsymbol{\tau}_{1'} = -\boldsymbol{\tau}_1 + 2\mathbf{r}_{1c} - \frac{2}{3}\mathbf{a}_1 - \frac{1}{3}\mathbf{a}_2 \tag{16}$$

$$\boldsymbol{\tau}_{2'} = -\boldsymbol{\tau}_2 + 2\mathbf{r}_{1c} + \frac{1}{3}\mathbf{a}_1 - \frac{1}{3}\mathbf{a}_2$$

$$\boldsymbol{\tau}_{3'} = -\boldsymbol{\tau}_3 + 2\mathbf{r}_{1c} + \frac{1}{3}\mathbf{a}_1 + \frac{2}{3}\mathbf{a}_2$$

where $\mathbf{r}_{1c}$ is the center of mass of sites 1, 2, 3. Note that if the kagome lattice is not distorted, we have

$$\boldsymbol{\tau}_1^0 = -\frac{1}{3}\mathbf{a}_1 - \frac{1}{6}\mathbf{a}_2, \ \ \boldsymbol{\tau}_2^0 = \frac{1}{6}\mathbf{a}_1 - \frac{1}{6}\mathbf{a}_2, \ \ \boldsymbol{\tau}_3^0 = \frac{1}{6}\mathbf{a}_1 + \frac{1}{3}\mathbf{a}_2 \tag{17}$$

In other words,

$$\boldsymbol{\tau}_{i'} = -\boldsymbol{\tau}_i + 2\boldsymbol{\tau}_i^0 + 2\mathbf{r}_{1c} \equiv \boldsymbol{\tau}_i^0 - \delta\boldsymbol{\tau}_i + 2\mathbf{r}_{1c}. \tag{18}$$

It is also easy to show that

$$\delta\boldsymbol{\tau}_i = -\delta\tau\hat{z} \times (\boldsymbol{\tau}_i^0 - \mathbf{r}_{1c}) \tag{19}$$

$$\delta\boldsymbol{\tau}_{i'} = \delta\tau\hat{z} \times (\boldsymbol{\tau}_i^0 - \mathbf{r}_{1c})$$

Naively, an arbitrary bond (together with its surrounding atoms or environment) between $(\mathbf{0}, i)$ and $(\mathbf{R}, j)$ will be transformed by $\mathcal{D}$ to that between $(\mathbf{0}, i')$ and $(\mathbf{R}, j')$. The two bond vectors are

$$\mathbf{r}_{ij}(\mathbf{R}) = \boldsymbol{\tau}_j - \boldsymbol{\tau}_i + \mathbf{R} = \boldsymbol{\tau}_j^0 - \boldsymbol{\tau}_i^0 + \mathbf{R} + (\delta\boldsymbol{\tau}_j - \delta\boldsymbol{\tau}_i) \tag{20}$$

$$\mathbf{r}_{i'j'}(\mathbf{R}) = \boldsymbol{\tau}_{j'} - \boldsymbol{\tau}_{i'} + \mathbf{R} = \boldsymbol{\tau}_j^0 - \boldsymbol{\tau}_i^0 + \mathbf{R} - (\delta\boldsymbol{\tau}_j - \delta\boldsymbol{\tau}_i)$$

To establish equivalence between the hopping parameters for the two systems, a first criterion is that the bond length remains unchanged. This leads to

$$(\delta\boldsymbol{\tau}_i - \delta\boldsymbol{\tau}_j) \cdot \mathbf{R} = 0 \tag{21}$$

which is a rather strong constraint. However, we can replace $\mathbf{r}_{i'j'}(\mathbf{R}')$ by $\mathbf{r}_{j'i'}(\mathbf{R}')$. Namely, the two sites are transformed to $(\mathbf{0}, j')$ and $(\mathbf{R}, i')$. Then instead of Eq. (21) we have

$$(\delta\boldsymbol{\tau}_j - \delta\boldsymbol{\tau}_i) \cdot (\mathbf{R} - \mathbf{R}') = 0. \tag{22}$$

Namely,

$$\mathbf{R}' = \mathcal{R}^{\pi}_{\boldsymbol{\tau}_j^0 + \boldsymbol{\tau}_i^0 - 2\mathbf{r}_{1c}}\mathbf{R}. \tag{23}$$

One can easily see that

$$\mathcal{R}^{\pi}_{\boldsymbol{\tau}_j^0 + \boldsymbol{\tau}_i^0 - 2\mathbf{r}_{1c}}\boldsymbol{\tau}_i = \boldsymbol{\tau}_{j'}, \ \ \mathcal{R}^{\pi}_{\boldsymbol{\tau}_j^0 + \boldsymbol{\tau}_i^0 - 2\mathbf{r}_{1c}}\boldsymbol{\tau}_j = \boldsymbol{\tau}_{i'}. \tag{24}$$

Therefore the hopping from $(\mathbf{0}, j')$ to $(\mathbf{R}', i')$ can be obtained from the hopping in the original lattice from $(\mathbf{0}, i)$ to $(\mathbf{R}, j)$ by performing the $\mathcal{R}^{\pi}_{\boldsymbol{\tau}_j^0 + \boldsymbol{\tau}_i^0 - 2\mathbf{r}_{1c}}$ transformation on the internal degrees of freedom of the model. Namely

$$T'_{\alpha\beta}(\mathbf{0}, j'; \mathbf{R}', i') = (U^{\dagger}TU)_{\alpha\beta}(\mathbf{0}, i; \mathbf{R}, j). \tag{25}$$

For the nearest neighbor bond in our model, because $T_{\alpha\beta}$ only has ↑↑ and ↓↓ elements, which change sign under $\mathcal{R}^{\pi}_{\boldsymbol{\tau}_j^0 + \boldsymbol{\tau}_i^0 - 2\mathbf{r}_{1c}}$, we have

$$T'_{\alpha\beta}(\mathbf{0}, j'; \mathbf{0}, i') = -T_{\alpha\beta}(\mathbf{0}, i; \mathbf{0}, j) = T_{\alpha\beta}(\mathbf{0}, j; \mathbf{0}, i). \tag{26}$$

Therefore the nearest-neighbor spin-dependent hopping is invariant under distortion reversal. In more complex models, however, Eq. (25) is the general relation. This can also be seen from Fig. S13 which shows the distortion-reversed lattice ($\mathcal{D}$ is equivalent to $\mathcal{R}^{\pi}_y$ in the paramagnetic state). But then $\mathcal{D}$ does not necessarily keep the nonmagnetic Hamiltonian invariant.

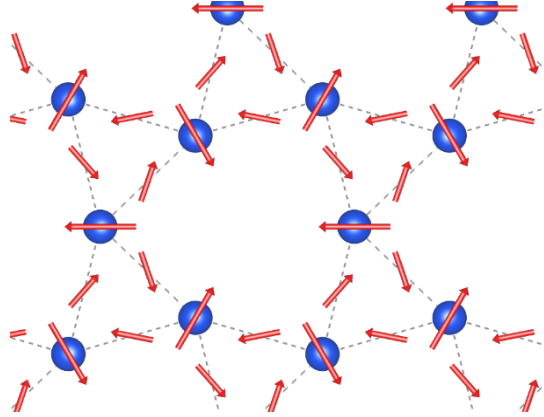

**Fig. S13| Nonmagnetic part of the tight-binding model transformed by distortion reversal.** The arrows represent the directions and sizes of the electric dipole moments defined in Eq. (11).

Up to now we have only considered in-plane hopping and the case of $i \neq j$. If $\mathbf{R} \cdot \hat{z} \neq 0$, it is not difficult to see that Eq. (25) still holds. However, for the nearest-neighbor out-of-plane hopping, we have

$$T'_{\alpha\beta}(\mathbf{0}, i'; c\hat{z}, i') = (U^\dagger T U)_{\alpha\beta}(\mathbf{0}, i; -c\hat{z}, i). \tag{27}$$

Since the $\hat{e}$ vectors on the inter-layer bonds in Fig. S12 are perpendicular to $\boldsymbol{\tau}_i^0 - \mathbf{r}_{1c}$, $(U^\dagger T U)$ is invariant under the $\pi$ rotation about $\boldsymbol{\tau}_i^0 - \mathbf{r}_{1c}$. Therefore

$$T'_{\alpha\beta}(\mathbf{0}, i'; c\hat{z}, i') = T_{\alpha\beta}(\mathbf{0}, i; -c\hat{z}, i). \tag{28}$$

This relation can also be seen by comparing Fig. S13 with Fig. S12.

Collecting all terms, we found that if given the original mean-field Hamiltonian with nearest-neighbor hopping only

$$
\begin{aligned}
H = {} & \sum_{\langle i, \mathbf{R}; j, \mathbf{R}' \rangle; \alpha} t(i, \mathbf{0}; j, \mathbf{R}' - \mathbf{R}) c^\dagger_{\alpha, i, \mathbf{R}} c_{\alpha, j, \mathbf{R}'} \\
& + \sum_{\langle i, \mathbf{R}; j, \mathbf{R}' \rangle; \alpha\beta} \lambda_{\alpha\beta}(i, \mathbf{0}; j, \mathbf{R}' - \mathbf{R}) c^\dagger_{\alpha, i, \mathbf{R}} c_{\beta, j, \mathbf{R}'} \\
& + \sum_{i, \mathbf{R}; \alpha\beta} \mathbf{J}_{i, \mathbf{R}} \cdot \boldsymbol{\sigma}_{\alpha\beta} c^\dagger_{\alpha, i, \mathbf{R}} c_{\beta, i, \mathbf{R}}
\end{aligned}
\tag{29}
$$

where the first term involves spin-independent hopping only, the second term includes nearest-neighbor spin-dependent hopping, and the lattice vectors are that of the nonmagnetic lattice, the Hamiltonian of the distortion-reversed system is

$$
\begin{aligned}
H' = {} & \sum_{\langle i, \mathbf{R}; j, \mathbf{R}' \rangle; \alpha} t(i, \mathbf{0}; j, (\mathbf{R}' - \mathbf{R})_\parallel - (\mathbf{R}' - \mathbf{R})_z) d^\dagger_{\alpha, i, \mathbf{R}} d_{\alpha, j, \mathbf{R}'} \\
& + \sum_{\langle i, \mathbf{R}; j, \mathbf{R}' \rangle; \alpha\beta} \lambda_{\alpha\beta}(i, \mathbf{0}; j, (\mathbf{R}' - \mathbf{R})_\parallel - (\mathbf{R}' - \mathbf{R})_z) d^\dagger_{\alpha, i, \mathbf{R}} d_{\beta, j, \mathbf{R}'} \\
& + \sum_{i, \mathbf{R}; \alpha\beta} \mathbf{J}_{i, \mathbf{R}} \cdot \boldsymbol{\sigma}_{\alpha\beta} d^\dagger_{\alpha, i, \mathbf{R}} d_{\beta, i, \mathbf{R}}.
\end{aligned}
\tag{30}
$$

After Fourier transform, the two Hamiltonians in momentum space are therefore related by $H(\mathbf{k}_\parallel, k_z) = H'(\mathbf{k}_\parallel, -k_z)$, and have identical band energies at the same $k_{x,y}$ but opposite $k_z$. The $\mathcal{R}_y^\pi$ operation following $\mathcal{D}$ brings $(k_x, k_y, -k_z)$ further to $(-k_x, k_y, k_z)$, which due to a $\mathcal{R}_x^\pi \mathcal{I}$ symmetry in Am'm2' is equivalent to $(k_x, k_y, k_z)$. This is the reason for the exactly identical band structures of $S_{1/3}$ and $S'_{1/3}$.

Although the Hamiltonian matrix is invariant under the distortion reversal (up to a sign change of the $z$ coordinate), the velocity operator matrix elements are different. This is because the position operator in the Wannier basis has the form

$$\mathbf{r} = \sum_{ab} |a\rangle\langle a|\mathbf{r}|b\rangle\langle b| = \sum_{ab} |a'\rangle\langle a'|\mathbf{r}|b'\rangle\langle b'| \neq \sum_{ab} |a'\rangle\langle a|\mathbf{r}|b\rangle\langle b'| \tag{31}$$

If using Wannier functions based on exponentially localized, atomic-orbital-like wave functions, one can approximately ignore the inter-orbital matrix elements of the position operator. Then the relation above becomes

$$\mathbf{r} \approx \sum_a |a\rangle \mathbf{r}_a \langle a| = \sum_a |a'\rangle \mathbf{r}_{a'} \langle a'| \neq \sum_a |a'\rangle \mathbf{r}_a \langle a'|. \tag{32}$$

In particular, the one-body velocity operator of the original system is

$$\mathbf{v} = \frac{1}{i\hbar} [\mathbf{r}, H] = \frac{1}{i\hbar} \sum_{ab} |a\rangle (\mathbf{r}_a - \mathbf{r}_b) H_{ab} \langle b| \tag{33}$$

while it becomes in the distortion-reversed system (for now we consider the case with $\langle a'|H'|b'\rangle = \langle a|H|b\rangle$, and discuss the additional effect of $z \to -z$ in $\mathcal{D}$ later)

$$\mathbf{v}' = \frac{1}{i\hbar} [\mathbf{r}, H'] = \frac{1}{i\hbar} \sum_{ab} |a'\rangle (\mathbf{r}_{a'} - \mathbf{r}_{b'}) H_{ab} \langle b'| \tag{34}$$

$$= \frac{1}{i\hbar} \sum_{ab} |a'\rangle (\mathbf{r}_a - \mathbf{r}_b) H_{ab} \langle b'| + \frac{1}{i\hbar} \sum_{ab} |a'\rangle [(\mathbf{r}_{a'} - \mathbf{r}_a) - (\mathbf{r}_{b'} - \mathbf{r}_b)] H_{ab} \langle b'|$$

$$\equiv U^\dagger \mathbf{v} U + U^\dagger \delta \mathbf{v} U$$

where the matrix $U$ is the basis transformation between $\{|a\rangle\}$ and $\{|a'\rangle\}$. Since the velocity or equivalently the current operator has an extra term besides the unitary transformation of $\mathbf{v}$, linear response functions involving the current operator will in general be different. In the case of intrinsic anomalous Hall conductivity, we have for the original system

$$\boldsymbol{\sigma}_{\text{AH}} = \frac{e^2 \hbar}{2} \sum_{m \neq n} \int \frac{d^3\mathbf{k}}{(2\pi)^3} \frac{f_n - f_m}{(\epsilon_n - \epsilon_m)^2} \text{Im}(\mathbf{v}_{nm} \times \mathbf{v}_{mn}) \tag{35}$$

where $m, n$ label bands. For the distortion-reversed system, we have

$$\boldsymbol{\sigma}'_{\text{AH}} = \frac{e^2 \hbar}{2} \sum_{m \neq n} \int \frac{d^3\mathbf{k}}{(2\pi)^3} \frac{f_n - f_m}{(\epsilon_n - \epsilon_m)^2} \text{Im}(\mathbf{v}'_{nm} \times \mathbf{v}'_{mn}) \tag{36}$$

$$= \boldsymbol{\sigma}_{\text{AH}} + e^2 \hbar \sum_{m \neq n} \int \frac{d^3\mathbf{k}}{(2\pi)^3} \frac{f_n - f_m}{(\epsilon_n - \epsilon_m)^2} \text{Im}(\mathbf{v}_{nm} \times \delta\mathbf{v}_{mn})$$

$$+ \frac{e^2 \hbar}{2} \sum_{m \neq n} \int \frac{d^3\mathbf{k}}{(2\pi)^3} \frac{f_n - f_m}{(\epsilon_n - \epsilon_m)^2} \text{Im}(\delta\mathbf{v}_{nm} \times \delta\mathbf{v}_{mn})$$

To get the difference between $\boldsymbol{\sigma}'_{\text{AH}}$ and $\boldsymbol{\sigma}_{\text{AH}}$, we need the explicit form of $\delta\mathbf{v}_{mn}$. This can be done by noticing that $\mathbf{r}$ in the basis $\{|a\rangle\} = \{|\alpha, i, \mathbf{R}\rangle\}$ is

$$\mathbf{r} = \sum_{\alpha, i, \mathbf{R}} |\alpha, i, \mathbf{R}\rangle (\boldsymbol{\tau}_i + \mathbf{R}) \langle \alpha, i, \mathbf{R}|, \tag{37}$$

which leads to the momentum space representation

$$\mathbf{r} = \sum_{\alpha, i, \mathbf{k}, \mathbf{k}'} |\alpha, i, \mathbf{k}\rangle [\boldsymbol{\tau}_i \delta_{\mathbf{k}, \mathbf{k}'} + \sum_{\mathbf{R}} \mathbf{R} e^{i\mathbf{R} \cdot (\mathbf{k}' - \mathbf{k})}] \langle \alpha, i, \mathbf{k}'|. \tag{38}$$

The relationship between Wannier and Bloch states is

$$|\alpha, i, \mathbf{k}\rangle = \sum_{\mathbf{R}} e^{i\mathbf{k} \cdot \mathbf{R}} |\alpha, i, \mathbf{R}\rangle, \tag{39}$$

$$|\alpha, i, \mathbf{R}\rangle = \frac{1}{N} \sum_{\mathbf{k} \in BZ} e^{-i\mathbf{k} \cdot \mathbf{R}} |\alpha, i, \mathbf{k}\rangle = \frac{1}{\Omega_{\text{BZ}}} \int_{\text{BZ}} d^3\mathbf{k} e^{-i\mathbf{k} \cdot \mathbf{R}} |\alpha, i, \mathbf{k}\rangle.$$

On the other hand, the Hamiltonian $H$ in this basis is

$$H = \sum_{\alpha,i,\mathbf{k};\beta,j,\mathbf{k}'} |\alpha,i,\mathbf{k}\rangle \left[ H_{\alpha i,\beta j}(\mathbf{k})\delta_{\mathbf{k},\mathbf{k}'} \right] \langle \beta,j,\mathbf{k}'| \tag{40}$$

where

$$H_{\alpha i,\beta j}(\mathbf{k}) = \sum_{\mathbf{R}} e^{i\mathbf{k}\cdot\mathbf{R}} \langle \alpha,i,\mathbf{0}|H|\beta,j,\mathbf{R}\rangle = \sum_{\mathbf{R}} e^{i\mathbf{k}\cdot\mathbf{R}} t_{\alpha i,\beta j}(\mathbf{R}). \tag{41}$$

We can then obtain the velocity operator

$$\mathbf{v} = \sum_{\alpha,i;\beta,j;\mathbf{k}} |\alpha,i,\mathbf{k}\rangle \left[ \frac{1}{\hbar}\nabla_{\mathbf{k}} H_{\alpha i,\beta j}(\mathbf{k}) + \frac{1}{i\hbar}(\boldsymbol{\tau}_i - \boldsymbol{\tau}_j) H_{\alpha i,\beta j}(\mathbf{k}) \right] \langle \beta,j,\mathbf{k}|. \tag{42}$$

Therefore $\delta\mathbf{v}$ is (note that $\delta\boldsymbol{\tau}_i \equiv \boldsymbol{\tau}_i - \boldsymbol{\tau}_0 = -(\boldsymbol{\tau}_i' - \boldsymbol{\tau}_0)$)

$$\delta\mathbf{v} = \frac{1}{i\hbar}\sum_{\alpha,i;\beta,j;\mathbf{k}} |\alpha,i,\mathbf{k}\rangle \left[ -2(\delta\boldsymbol{\tau}_i - \delta\boldsymbol{\tau}_j) H_{\alpha i,\beta j}(\mathbf{k}) \right] \langle \beta,j,\mathbf{k}| \tag{43}$$

which can be understood as the commutator between a Berry connection matrix defined as

$$\mathcal{A}_{\alpha i,\beta j}(\mathbf{k}) = -2\delta\boldsymbol{\tau}_i \delta_{ij}\delta_{\alpha\beta} \tag{44}$$

and the momentum-space Hamiltonian.

It is sometimes convenient to include a phase factor in the definition of Bloch states in terms of Wannier states:

$$|\alpha,i,\mathbf{k}\rangle_g = e^{i\mathbf{k}\cdot\boldsymbol{\tau}_i}\sum_{\mathbf{R}} e^{i\mathbf{k}\cdot\mathbf{R}}|\alpha,i,\mathbf{R}\rangle = e^{i\mathbf{k}\cdot\boldsymbol{\tau}_i}|\alpha,i,\mathbf{k}\rangle, \tag{45}$$

$$|\alpha,i,\mathbf{R}\rangle = \frac{1}{N}\sum_{\mathbf{k}\in BZ} e^{-i\mathbf{k}\cdot\boldsymbol{\tau}_i}e^{-i\mathbf{k}\cdot\mathbf{R}}|\alpha,i,\mathbf{k}\rangle_g = \frac{1}{\Omega_{BZ}}\int_{BZ} d^3k\, e^{-i\mathbf{k}\cdot(\boldsymbol{\tau}_i+\mathbf{R})}|\alpha,i,\mathbf{k}\rangle_g.$$

Such a phase or gauge choice has certain benefits. For example, the momentum space Hamiltonian in this basis becomes

$$H_{\alpha i,\beta j}^g(\mathbf{k}) = \sum_{\mathbf{R}} e^{i\mathbf{k}\cdot(\mathbf{R}+\boldsymbol{\tau}_j-\boldsymbol{\tau}_i)} \langle \alpha,i,\mathbf{0}|H|\beta,j,\mathbf{R}\rangle = \sum_{\mathbf{R}} e^{i\mathbf{k}\cdot(\mathbf{R}+\boldsymbol{\tau}_j-\boldsymbol{\tau}_i)} t_{\alpha i,\beta j}(\mathbf{R}) \tag{46}$$

$$= H_{\alpha i,\beta j}(\mathbf{k})e^{i\mathbf{k}\cdot(\boldsymbol{\tau}_j-\boldsymbol{\tau}_i)}$$

so that the phase factor only depends on the position vectors connecting two neighboring sites. Moreover, the matrix elements of the velocity operator can be written as

$$\mathbf{v}_{\alpha i,\beta j}(\mathbf{k}) = \frac{1}{\hbar}\nabla_{\mathbf{k}} H_{\alpha i,\beta j}(\mathbf{k}) + \frac{1}{i\hbar}(\boldsymbol{\tau}_i - \boldsymbol{\tau}_j) H_{\alpha i,\beta j}(\mathbf{k}) \tag{47}$$

$$= \frac{1}{\hbar}\nabla_{\mathbf{k}}[H_{\alpha i,\beta j}^g(\mathbf{k})e^{-i\mathbf{k}\cdot(\boldsymbol{\tau}_j-\boldsymbol{\tau}_i)}] + \frac{1}{i\hbar}(\boldsymbol{\tau}_i - \boldsymbol{\tau}_j)H_{\alpha i,\beta j}^g(\mathbf{k})e^{-i\mathbf{k}\cdot(\boldsymbol{\tau}_j-\boldsymbol{\tau}_i)}$$

$$= \frac{1}{\hbar}[\nabla_{\mathbf{k}}H_{\alpha i,\beta j}^g(\mathbf{k})]e^{-i\mathbf{k}\cdot(\boldsymbol{\tau}_j-\boldsymbol{\tau}_i)}.$$

As a result

$$\mathbf{v} = \sum_{\alpha,i;\beta,j;\mathbf{k}} e^{i\mathbf{k}\cdot\boldsymbol{\tau}_i}|\alpha,i,\mathbf{k}\rangle \left[ \frac{1}{\hbar}\nabla_{\mathbf{k}}H_{\alpha i,\beta j}^g(\mathbf{k}) \right] \langle \beta,j,\mathbf{k}|e^{-i\mathbf{k}\cdot\boldsymbol{\tau}_j} \tag{48}$$

$$= \sum_{\alpha,i;\beta,j;\mathbf{k}} |\alpha, i, \mathbf{k}\rangle_g \left[\frac{1}{\hbar} \nabla_{\mathbf{k}} H^g_{\alpha i,\beta j}(\mathbf{k})\right] \langle \beta, j, \mathbf{k}|_g.$$

On the other hand, $\mathbf{v}'$ is

$$\mathbf{v}' = \sum_{\alpha,i;\beta,j;\mathbf{k}} |\alpha, i, \mathbf{k}\rangle'_g \left[\frac{1}{\hbar} \nabla_{\mathbf{k}} H'^g_{\alpha i,\beta j}(\mathbf{k})\right] \langle \beta, j, \mathbf{k}|'_g. \tag{49}$$

Although the above two expressions are similar, they have different matrix elements in their respective basis because $\nabla_{\mathbf{k}} H^g_{\alpha i,\beta j}(\mathbf{k}) \neq \nabla_{\mathbf{k}} H'^g_{\alpha i,\beta j}(\mathbf{k})$. The $\delta \mathbf{v}$ (or rather $U^\dagger \delta \mathbf{v} U$) in this gauge choice is

$$U^\dagger \delta \mathbf{v} U = \sum_{\alpha,i;\beta,j;\mathbf{k}} |\alpha, i, \mathbf{k}\rangle'_g \, \delta \mathbf{v}^g_{\alpha i,\beta j}(\mathbf{k}) \langle \beta, j, \mathbf{k}|'_g. \tag{50}$$

where

$$\delta \mathbf{v}^g_{\alpha i,\beta j}(\mathbf{k}) = \frac{1}{\hbar}\left[e^{-i\mathbf{k}\cdot(\boldsymbol{\tau}_{j\prime}-\boldsymbol{\tau}_{i\prime})}\nabla_{\mathbf{k}} H'^g_{\alpha i,\beta j}(\mathbf{k}) - e^{-i\mathbf{k}\cdot(\boldsymbol{\tau}_j-\boldsymbol{\tau}_i)}\nabla_{\mathbf{k}} H^g_{\alpha i,\beta j}(\mathbf{k})\right] e^{i\mathbf{k}\cdot(\boldsymbol{\tau}_{j\prime}-\boldsymbol{\tau}_{i\prime})}$$

$$= \frac{1}{\hbar}\nabla_{\mathbf{k}}\left[H^g_{\alpha i,\beta j}(\mathbf{k})e^{-i\mathbf{k}\cdot(2\delta\boldsymbol{\tau}_j-2\delta\boldsymbol{\tau}_i)}\right] - \frac{1}{\hbar}e^{-i\mathbf{k}\cdot(2\delta\boldsymbol{\tau}_j-2\delta\boldsymbol{\tau}_i)}\nabla_{\mathbf{k}} H^g_{\alpha i,\beta j}(\mathbf{k})$$

$$= \frac{2i}{\hbar}(\delta\boldsymbol{\tau}_i - \delta\boldsymbol{\tau}_j)H^g_{\alpha i,\beta j}(\mathbf{k})e^{-i\mathbf{k}\cdot(2\delta\boldsymbol{\tau}_j-2\delta\boldsymbol{\tau}_i)} \tag{51}$$

in agreement with Eq. (43).

The derivation above only considers the situation that $\langle a'|H'|b'\rangle = \langle a|H|b\rangle$. If we specialize to the case that $\mathcal{D}$ also changes the Hamiltonian matrix by $k_z \to -k_z$, the z component of the velocity operator $v_z$ will have an additional sign change. This leads to the largely opposite Berry curvature and $\boldsymbol{\sigma}_{\mathrm{AH}}$ between $S_{1/3}$ and $S'_{1/3}$. It also makes the sum rather than the difference of the $\boldsymbol{\sigma}_{\mathrm{AH}}$ of the two states correspond to the contributions by $\mathcal{A}$, which is the reason why the mean of the two curves in Fig. 4d (or 4f for the 2/3 plateau case) has a weak dependence on Fermi energy.

In general, for a magnetic crystal that has certain distortion, and the distortion-reversed system with the same magnetic order has nearly the same energy due to some emergent distortion reversal symmetry and/or weak spin-orbit coupling, the two systems can in principle have very different anomalous Hall effect since the distortion reversal generally changes the current operator. Conversely, if two magnetic states can be made equivalent through a proper rotation together with a distortion reversal, while their energies may or may not be the same depending on the underlying symmetry of the system, the AHE of the two magnetic states should in general have different sizes.

## Supplementary Note 6: Transformation of the classical spin model under distortion reversal

In this section we discuss why the classical spin model of HoAgGe also gives identical energies for $S_{1/3}$ and $S'_{1/3}$. The classical spin model is

$$H = \frac{1}{2}\sum_{ij} J_{ij}\hat{n}_i \cdot \hat{n}_j - \frac{1}{2}\sum_{ij} \frac{\mu_0 g^2 \mu_B^2}{4\pi r_{ij}^3}\left[3(\hat{n}_i \cdot \hat{r}_{ij})(\hat{n}_j \cdot \hat{r}_{ij}) - \hat{n}_i \cdot \hat{n}_j\right] \quad (52)$$

where $\hat{n}_i$ is a unit vector along the spin direction on site $i$, and $i, j$ label both sublattices and unit cells.

We start with the dipolar term. Since the classical spin Hamiltonian is pairwise we can consider an arbitrary term in the sum and recover the sublattice and unit cell indices:

$$-\frac{\mu_0 g^2 \mu_B^2}{4\pi r_{i,\mathbf{R};j,\mathbf{R}'}^3}\left[3(\hat{n}_{i,\mathbf{R}} \cdot \hat{r}_{i,\mathbf{R};j,\mathbf{R}'})(\hat{n}_{j,\mathbf{R}'} \cdot \hat{r}_{i,\mathbf{R};j,\mathbf{R}'}) - \hat{n}_{i,\mathbf{R}} \cdot \hat{n}_{j,\mathbf{R}'}\right]. \quad (53)$$

Distortion reversal only changes $\mathbf{r}_{i,\mathbf{R};j,\mathbf{R}'}$. For $i = j$ it is trivial to prove that the above term does not change. Therefore we only consider $i \neq j$ below. Moreover, we define

$$\hat{n}_{i,\mathbf{R}} = \sigma_{i,\mathbf{R}}\hat{n}_i \quad (54)$$

where $\sigma_{i,\mathbf{R}} = \pm 1$ determines the direction of the Ising spins for a given magnetic state, and $\hat{n}_i$ is the direction of the local easy axis. Before the reversal we have

$$\mathbf{r}_{i,\mathbf{R};j,\mathbf{R}'} = (\mathbf{R}' - \mathbf{R}) + \boldsymbol{\tau}_j^0 - \boldsymbol{\tau}_i^0 + (\delta\boldsymbol{\tau}_j - \delta\boldsymbol{\tau}_i) \quad (55)$$

and after the reversal it becomes

$$\mathbf{r}_{i',\mathbf{R};j',\mathbf{R}'} = (\mathbf{R}' - \mathbf{R}) + \boldsymbol{\tau}_j^0 - \boldsymbol{\tau}_i^0 - (\delta\boldsymbol{\tau}_j - \delta\boldsymbol{\tau}_i). \quad (56)$$

In general $r_{i',\mathbf{R};j',\mathbf{R}'} \neq r_{i,\mathbf{R};j,\mathbf{R}'}$. However, in the previous section we found that one can consider the pair in the distortion-reversed lattice $(j', \widetilde{\mathbf{R}}) - (i', \widetilde{\mathbf{R}}')$, which are related to $(i, \mathbf{R})$ and $(j, \mathbf{R}')$ by

$$\boldsymbol{\tau}_{j'} + \widetilde{\mathbf{R}} = \mathcal{R}^\pi_{\boldsymbol{\tau}_j^0 + \boldsymbol{\tau}_i^0 - 2\mathbf{r}_{1c}}(\boldsymbol{\tau}_i + \mathbf{R}), \ \ \boldsymbol{\tau}_{i'} + \widetilde{\mathbf{R}}' = \mathcal{R}^\pi_{\boldsymbol{\tau}_j^0 + \boldsymbol{\tau}_i^0 - 2\mathbf{r}_{1c}}(\boldsymbol{\tau}_j + \mathbf{R}') \quad (57)$$

and has the property $r_{j',\widetilde{\mathbf{R}};i',\widetilde{\mathbf{R}}'} = r_{i,\mathbf{R};j,\mathbf{R}'}$. Moreover

$$\hat{n}_{j',\widetilde{\mathbf{R}}} \cdot \hat{r}_{j',\widetilde{\mathbf{R}};i',\widetilde{\mathbf{R}}'} = \sigma_{j',\widetilde{\mathbf{R}}}\hat{n}_j \cdot \hat{r}_{j',\widetilde{\mathbf{R}};i',\widetilde{\mathbf{R}}'} = \sigma_{j',\widetilde{\mathbf{R}}}(\mathcal{R}^\pi_{\boldsymbol{\tau}_j^0 + \boldsymbol{\tau}_i^0 - 2\mathbf{r}_{1c}}\hat{n}_j) \cdot \hat{r}_{i,\mathbf{R};j,\mathbf{R}'} \quad (58)$$

$$= \sigma_{j',\widetilde{\mathbf{R}}}\hat{n}_i \cdot \hat{r}_{i,\mathbf{R};j,\mathbf{R}'}.$$

Therefore

$$(\hat{n}_{j',\widetilde{\mathbf{R}}} \cdot \hat{r}_{j',\widetilde{\mathbf{R}};i',\widetilde{\mathbf{R}}'})(\hat{n}_{i',\widetilde{\mathbf{R}}'} \cdot \hat{r}_{j',\widetilde{\mathbf{R}};i',\widetilde{\mathbf{R}}'}) = \sigma_{j',\widetilde{\mathbf{R}}}\sigma_{i',\widetilde{\mathbf{R}}'}(\hat{n}_i \cdot \hat{r}_{i,\mathbf{R};j,\mathbf{R}'})(\hat{n}_j \cdot \hat{r}_{i,\mathbf{R};j,\mathbf{R}'}). \quad (59)$$

In other words, if $\sigma_{j',\tilde{\mathbf{R}}} = \sigma_{j,\mathbf{R}'}$ and $\sigma_{i',\tilde{\mathbf{R}}'} = \sigma_{i,\mathbf{R}}$, the dipolar interaction will be invariant. This requires both $\tilde{\mathbf{R}} - \mathbf{R}'$ and $\tilde{\mathbf{R}}' - \mathbf{R}$ to be certain magnetic lattice vectors. Since the unit lattice vector along z does not change in the magnetically ordered states, we only consider 2D lattice vectors. Then

$$\tilde{\mathbf{R}} - \mathbf{R}' = \mathcal{R}^{\pi}_{\boldsymbol{\tau}_j^0 + \boldsymbol{\tau}_i^0 - 2\mathbf{r}_{1c}} \mathbf{R} - \mathbf{R}' = -(2\mathbf{R} \cdot \hat{t}_{ij}^0)\hat{t}_{ij}^0 + (\mathbf{R} - \mathbf{R}') \tag{60}$$

where $\hat{t}_{ij}^0 \equiv (\boldsymbol{\tau}_j^0 - \boldsymbol{\tau}_i^0)/|\boldsymbol{\tau}_j^0 - \boldsymbol{\tau}_i^0|$. Since $\tau_{ij}^0$ is along a unit lattice vector of the nonmagnetic lattice, without loss of generality we can assume

$$\mathbf{R} - \mathbf{R}' = m\mathbf{a}_1 + n\mathbf{a}_2, \ (2\mathbf{R} \cdot \hat{t}_{ij}^0)\hat{t}_{ij}^0 = (2p - q)\mathbf{a}_1. \tag{61}$$

where $m, n, p, q$ are integers. $m, n$ are fixed for a given pair of sites, but $p, q$ can change freely depending on the choice of origin or the position of the mirror plane. Therefore

$$\tilde{\mathbf{R}} - \mathbf{R}' = (m - 2p + q)\mathbf{a}_1 + n\mathbf{a}_2 \tag{62}$$

The $\sqrt{3} \times \sqrt{3}$ unit cell has the following lattice vectors

$$\mathbf{b}_1 = 2\mathbf{a}_1 + \mathbf{a}_2, \ \mathbf{b}_2 = \mathbf{a}_1 + 2\mathbf{a}_2. \tag{63}$$

Therefore a general magnetic lattice vector is

$$\mathbf{R}^M = l\mathbf{b}_1 + r\mathbf{b}_2 = (2l + r)\mathbf{a}_1 + (l + 2r)\mathbf{a}_2. \tag{64}$$

By requiring $\tilde{\mathbf{R}} - \mathbf{R}' = \mathbf{R}^M$ we get

$$l + 2r = n, \ 2l + r = m - 2p + q, \tag{65}$$

which gives

$$r = \frac{2n - m + 2p - q}{3}, \ l = n - 2r. \tag{66}$$

$r, l$ can be integers if we let, e.g. $p = -n, q = -m$. On the other hand,

$$\tilde{\mathbf{R}}' - \mathbf{R} = -(2\mathbf{R}' \cdot \hat{t}_{ij}^0)\hat{t}_{ij}^0 - (\mathbf{R} - \mathbf{R}') \tag{67}$$

$$= -2[(\mathbf{R}' - \mathbf{R} + \mathbf{R}) \cdot \hat{t}_{ij}^0]\hat{t}_{ij}^0 - (\mathbf{R} - \mathbf{R}')$$

$$= (m - n - 2p + q)\mathbf{a}_1 - n\mathbf{a}_2$$

By requiring that $\tilde{\mathbf{R}}' - \mathbf{R} = \mathbf{R}^M$ we have

$$l + 2r = -n, \ 2l + r = m - n - 2p + q, \tag{68}$$

which gives

$$r = \frac{-n - m + 2p - q}{3}, \ l = -n - 2r \tag{69}$$

Using the above values, i.e. $p = -n, q = -m$ we can get $r = -n$ and $l = n$. Therefore both conditions can be satisfied simultaneously.

Moreover, the second term in the dipolar interaction is also invariant under distortion reversal, since

$$\hat{n}_{i,\mathbf{R}} \cdot \hat{n}_{j,\mathbf{R}'} = \sigma_{i,\mathbf{R}}\sigma_{j,\mathbf{R}'}\hat{n}_i \cdot \hat{n}_j = \sigma_{i',\widetilde{\mathbf{R}}'}\sigma_{j',\widetilde{\mathbf{R}}}\hat{n}_i \cdot \hat{n}_j = \hat{n}_{j',\widetilde{\mathbf{R}}} \cdot \hat{n}_{i',\widetilde{\mathbf{R}}'} \tag{70}$$

We have thus proved that for any two sites $(i, \mathbf{R})$ and $(j, \mathbf{R}')$, there exists another pair $(i', \widetilde{\mathbf{R}}')$ and $(j', \widetilde{\mathbf{R}})$ in the distortion-reversed lattice that has the same dipolar interaction. Therefore the total dipolar interaction energy is invariant under distortion reversal.

Finally we consider the Heisenberg term. Using the same procedure as above, we can see that

$$J_{i,\mathbf{R};j,\mathbf{R}'}\hat{n}_{i,\mathbf{R}} \cdot \hat{n}_{j,\mathbf{R}'} = J_{i,\mathbf{R};j,\mathbf{R}'}\hat{n}_{j',\widetilde{\mathbf{R}}} \cdot \hat{n}_{i',\widetilde{\mathbf{R}}'}. \tag{71}$$

If the Heisenberg exchange coupling is only determined by the distance between two spins, which is the case in the present model, we have $J_{i,\mathbf{R};j,\mathbf{R}'} = J_{j',\widetilde{\mathbf{R}};i',\widetilde{\mathbf{R}}'}$. Therefore each term in the Heisenberg interaction of the original system is equal to a counterpart in that of the distortion-reversed system. The total Heisenberg interaction energy of the classical spin model is invariant under distortion reversal regardless of the number of neighbors taken into account.

## Supplementary Note 7: Degenerate states with nonzero out-of-plane spin components

Our thermodynamic and neutron experimental data indicate that the Ho spins have finite out-of-plane components which become ordered at around 2.5 K. The ordering of the out-of-plane components does not further change the magnetic unit cell for the ground state and the plateau states based on the neutron data. Moreover, the coupling between the out-of-plane spin components is predominantly antiferromagnetic as indicated by magnetic susceptibility data. While a complete determinantion of the low-temperature magnetic order and spin Hamiltonian is under current investigation and will be reported elsewhere, based on the available evidence we propose a couple of plausible magnetic structures of the out-of-plane spin components that can induce the time-reversal-like degeneracy for the 2/3 plateau while keeping the conclusions for the 1/3 plateau intact.

Motivated by the known results of antiferromagnetically coupled Ising spins on the kagome lattice we consider the following order (Fig. S14) [4], which is equivalent to the $\sqrt{3} \times \sqrt{3}$ ground state of the dipolar kagome ice when the magnetic moments are represented by Ising variables. The state is degenerate with 5 other states obtained by time-reversal and translation, equivalent to the 6-fold degeneracy of the $\sqrt{3} \times \sqrt{3}$ ground state of the kagome ice. Since the ordering of the in-plane components on the 2/3 plateau also has a six-fold degeneracy by time reversal and translation, there are 6 possible variants of each unique state of the in-plane components with $C2'$ symmetry. Moreover, these 6 variants can be grouped into 3 pairs, with the two states in each pair connected by the $\mathcal{R}_y^{\pi}\mathcal{D}$ operation. Fig. 4e in the main text shows one of such pairs. It is possible for a pair to be selected as the lowest energy one by more realistic couplings among the Ho spins, which will be left for future investigation. We note that it is also possible to have states breaking the 3-fold rotation symmetry for the out-of-plane components on the 2/3 plateau, which we will not discuss here.

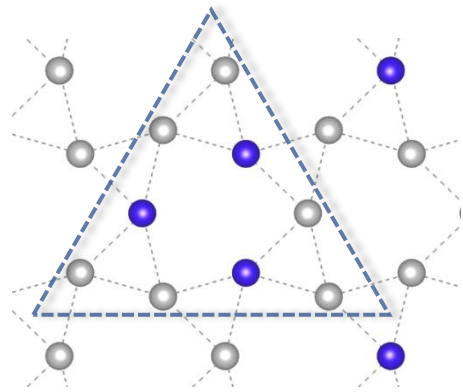

**Fig. S14| Possible order of the out-of-plane components of Ho spins on the 2/3 plateau.** Blue and gray spheres correspond to up and down directions, respectively. The triangular region bounded by the dashed lines includes the 9 sites in the magnetic unit cell. The depicted state is degenerate with 5 other states obtained by time reversal and translation.

Take the two states in Fig. 4e for example. They cannot be made equivalent by any operations in their symmetry group $C2'$, but can be transformed to each other by $\mathcal{R}_y^\pi \mathcal{D}$. Also since $\mathcal{R}_x^\pi \mathcal{I}$ is still a symmetry, the band structures of the two states will be identical, but the Berry curvatures will generally differ, as shown in the figure below.

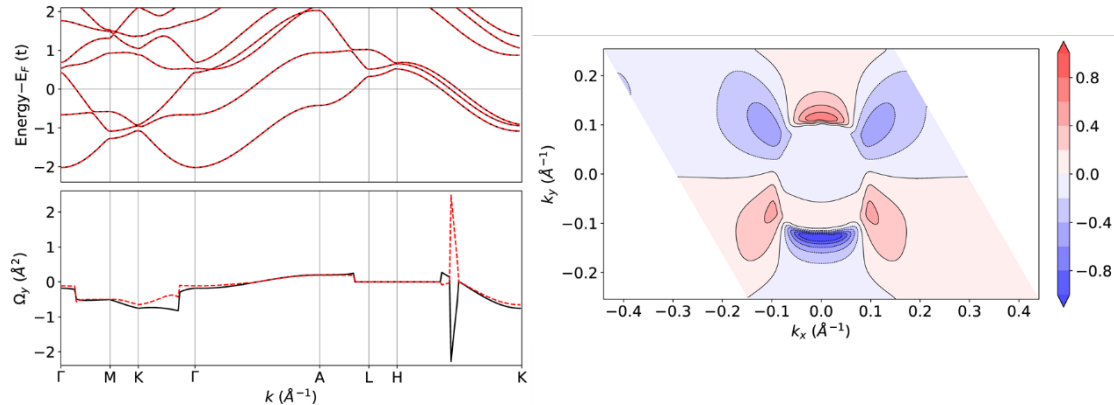

**Fig. S15| Band structures and Berry curvatures for the two degenerate states on the 2/3 plateau.** The right panel shows the difference between the Berry curvature along $y$ for the two states in the $k_z = 0$ plane. The size of the out-of-plane components of the Ho spins are taken to be 0.2, while the in-plane ones are unity. The other parameter values are $t_v = -0.4$, $\lambda_h = 0.3$, $\lambda_v = 0.2$, $J = -2.2$, $\mu = -4.4$, where $t_v, \lambda_h, \lambda_v$ are nearest-neighbor out-of-plane hopping, in-plane spin-orbit coupling, and out-of-plane spin-orbit coupling, respectively.

Assuming the same out-of-plane components order also exists on the 1/3 plateau, each of $S_{1/3}$ and $S'_{1/3}$ will have 6 variants, similar to the 2/3 plateau case. However, the in-total 12 states can also be grouped into 6 pairs, with the two states in each pair coming from $S_{1/3}$ and $S'_{1/3}$ separately and can be transformed to each other by $\mathcal{R}_y^\pi \mathcal{D}$ (up to a translation). Fig. S16 shows such an example. We note that these are just possible candidates for the time-reversal-like degenerate states on the 1/3 plateau when the out-of-plane components order is taken into account. Nonetheless, these examples suggest that the physics discussed in the main text is still operative when the out-of-plane components are ordered.

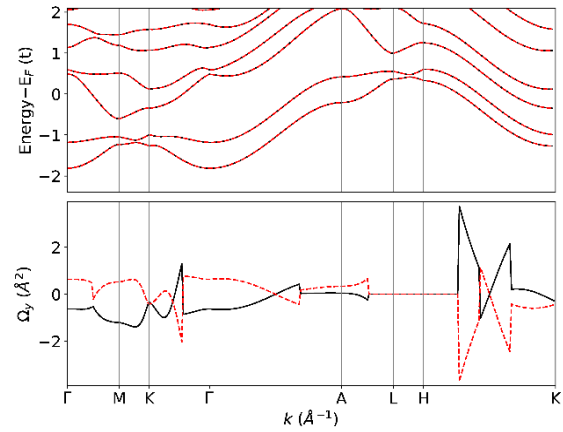

**Fig. S16| Band structures and Berry curvatures for two degenerate states on the 1/3 plateau with out-of-plane components order.** The parameter values are $t_v = -0.4$, $\lambda_h = 0.3$, $\lambda_v = 0.2$, $J = -2.2$, $\mu = -4.4$.